%% file: main_final.tex
\documentclass[10pt]{article} 
\usepackage[accepted]{tmlr}

\usepackage{acronym}
\input{math_commands.tex}

\input{ac.tex} 

\usepackage{url}
\usepackage{hyperref}
\usepackage{booktabs}
\usepackage{rotating}
\usepackage{tabularx}
\usepackage[edges]{forest}
\usepackage{xcolor}
\usepackage{graphicx}
\usepackage{makecell}
\usepackage{subcaption}
\usepackage{amssymb}
\usepackage{multirow}
\useforestlibrary{edges}

\hypersetup{
	colorlinks=true,
	citecolor=blue,
}

\definecolor{hidden-red}{RGB}{205, 44, 36}
\definecolor{hidden-blue}{RGB}{194,232,247}
\definecolor{hidden-orange}{RGB}{243,202,120}
\definecolor{hidden-green}{RGB}{34,139,34}
\definecolor{hidden-pink}{RGB}{255,245,247}
\definecolor{hidden-black}{RGB}{20,68,106}
\definecolor{purple}{RGB}{144,153,196}
\definecolor{yellow}{RGB}{255,228,123}
\definecolor{hidden-yellow}{RGB}{255,248,203}
\definecolor{tkcolor}{RGB}{224,223,255}
\definecolor{darkblue}{rgb}{0, 0.40, 0.75}

\definecolor{fixed}{RGB}{150, 0, 0}
\definecolor{multi}{RGB}{165, 90, 20}
\definecolor{agno}{RGB}{50, 3, 125}

\newcommand{\litbox}[1]{%
\begin{minipage}{25em} 
\raggedright\footnotesize 
#1
\end{minipage}%
}

\tikzset{
my-box/.style={
rectangle,
draw=black,
rounded corners,
line width=0.8pt,
text opacity=1,
minimum height=1.5em,
inner sep=2pt,
align=center,
fill opacity=.5
},
rootnode/.style={my-box,fill=black!15,font=\bfseries\small},
rate node/.style={my-box,fill=teal!25,font=\footnotesize},
fixed node/.style={my-box,fill=white,font=\footnotesize, text=fixed},
multi node/.style={my-box,fill=white,font=\footnotesize,text=multi},
agno node/.style={my-box,fill=white,font=\footnotesize, text=agno},
fixed leaf/.style={my-box,fill=hidden-red!20,align=left,inner xsep=2pt,inner ysep=2pt,yshift=-1pt},
multi leaf/.style={my-box,fill=yellow!32,align=left,inner xsep=2pt,inner ysep=2pt,yshift=-1pt},
agno leaf/.style={my-box,fill=darkblue!15,align=left,inner xsep=2pt,inner ysep=2pt,yshift=-1pt},
leaf node/.style={my-box,fill=white,align=left,inner xsep=2pt,inner ysep=2pt,yshift=-1pt}
}

\title{A Survey of Advancing Audio Super-Resolution and Bandwidth Extension from Discriminative to Generative Models}

\author{\name Ningyuan Yang~\thanks{Equal contribution. Link: \href{https://github.com/ningyuan33/Awesome-Audio-Super-Resolution-Bandwidth-Extension}{\textcolor{blue}{https://github.com/ningyuan33/Awesome-Audio-Super-Resolution-Bandwidth-Extension}}.} \email ningyuan.yang@stonybrook.edu \\
      \addr Department of Electrical and Computer Engineering\\
      Stony Brook University
      \AND
      \name Yize Li~\footnotemark[1] \email li.yize@northeastern.edu \\
      \addr Department of Electrical and Computer Engineering\\ Northeastern University
      \AND
      \name Diego A. Cuji \email diego.cujidutan@stonybrook.edu \\
      \addr Department of Electrical and Computer Engineering\\
      Stony Brook University
      \AND
      \name Ryan M. Corey \email corey1@uic.edu\\
      \addr Department of Electrical and Computer Engineering\\
      University of Illinois Chicago $\&$ Discovery Partners Institute
      \AND
      \name Pu Zhao \email p.zhao@northeastern.edu \\
      \addr Department of Electrical and Computer Engineering\\ Northeastern University
      \AND
      \name Xue Lin \email xue.lin@northeastern.edu \\
      \addr Department of Electrical and Computer Engineering\\ Northeastern University
      \AND
      \name Andrew C. Singer \email andrew.c.singer@stonybrook.edu\\
      \addr Department of Electrical and Computer Engineering\\
      Stony Brook University
      }

\def\month{11}  
\def\year{2026} 
\def\openreview{\url{https://openreview.net/forum?id=NdL2H5aoBn}} 

\begin{document}

\maketitle

\input{sec/0_abstract}
\input{sec/1_introduction_final}
\input{sec/2_background_final}
\input{sec/3_training}

\input{sec/4_evaluation}
\input{sec/5_discriminative}
\input{sec/6_generative_final}
\input{sec/7_quantitative_final}
\input{sec/8_discussion_final}

\input{sec/9_conclusion}
\input{sec/10_acknowledgement}



\bibliography{main}
\bibliographystyle{tmlr}

\appendix

\end{document}

%% file: math_commands.tex
\usepackage{amsmath,amsfonts,bm}

\def\eqref#1{equation~\ref{#1}}

\def\1{\bm{1}}

\def\mI{{\bm{I}}}

\DeclareMathAlphabet{\mathsfit}{\encodingdefault}{\sfdefault}{m}{sl}
\SetMathAlphabet{\mathsfit}{bold}{\encodingdefault}{\sfdefault}{bx}{n}



%% file: ac.tex
\newacro{dl}[DL]{deep learning}
\newacro{dnn}[DNN]{deep neural network}
\newacroplural{dnn}[DNNs]{deep neural networks}
\newacro{cnn}[CNN]{convolutional neural network}
\newacroplural{cnn}[CNNs]{convolutional neural networks}
\newacro{rnn}[RNN]{recurrent neural network}
\newacroplural{rnn}[RNNs]{recurrent neural networks}
\newacro{gan}[GAN]{generative adversarial network}
\newacroplural{gan}[GANs]{generative adversarial networks}
\newacro{vae}[VAE]{variational autoencoder}
\newacroplural{vae}[VAEs]{variational autoencoders}
\newacro{cvae}[CVAE]{conditional VAE}
\newacro{mlp}[MLP]{multilayer perceptron}
\newacroplural{mlp}[MLPs]{multilayer perceptrons}
\newacro{nb}[NB]{narrowband}
\newacro{wb}[WB]{wideband}
\newacro{bwe}[BWE]{bandwidth extension}
\newacro{ssr}[SSR]{speech super-resolution}
\newacro{asr}[SR]{audio super-resolution}
\newacro{se}[SE]{speech enhancement}
\newacro{lr}[LR]{low-resolution}
\newacro{hr}[HR]{high-resolution}
\newacro{lf}[LF]{low-frequency}
\newacro{hf}[HF]{high-frequency}
\newacro{stft}[STFT]{short-time Fourier transform}
\newacro{lps}[LPS]{log-power spectrum}
\newacro{mfcc}[MFCC]{mel-frequency cepstral coefficient}
\newacro{lsf}[LSF]{line spectral frequency}
\newacro{bpvc}[BPVC]{band-pass voicing coefficient}
\newacro{lpc}[LPC]{linear predictive coding}
\newacro{dct}[DCT]{discrete cosine transform}
\newacro{acf}[AFC]{autocorrelation function coefficient}
\newacro{zcr}[ZCR]{zero-crossing rate}
\newacro{gbrbm}[GBRBM]{Gaussian-Bernoulli restricted Boltzmann machine}
\newacro{pqmf}[PQMF]{Pseudo Quadrature Mirror Filter}
\newacro{mdct}[MDCT]{modified discrete cosine transform}
\newacro{hmm}[HMM]{hidden Markov model}
\newacroplural{hmm}[HMMs]{hidden Markov models}
\newacroplural{gmm}[GMMs]{Gaussian mixture models}
\newacro{ar}[AR]{autoregressive}
\newacro{llm}[LLM]{large language model}
\newacroplural{llm}[LLMs]{large language models}
\newacro{lalm}[LALM]{large audio-language model}
\newacroplural{lam}[LAMs]{large audio-language models}

\newacro{rvq}[RVQ]{residual vector quantization}
\newacro{ode}[ODE]{ordinary differential equation}
\newacroplural{CNF}[CNFs]{continuous normalizing flows}
\newacroplural{sde}[SDEs]{stochastic differential equations}
\newacro{snr}[SNR]{Signal-to-Noise Ratio}
\newacro{lsd}[LSD]{Log Spectral Distance}

\newacro{mmse}[MMSE]{minimum mean-square error}
\newacro{istft}[iSTFT]{inverse STFT}
\newacro{mse}[MSE]{mean squared error}
\newacro{mae}[MAE]{mean absolute error}
\newacro{sisdr}[SI-SDR]{scale-invariant signal-to-distortion ratio}
\newacro{nll}[NLL]{negative log-likelihood}
\newacro{ce}[CE]{cross-entropy}
\newacro{mrstft}[MR-STFT]{multi-resolution STFT}
\newacro{fm}[FM]{feature matching}
\newacro{lb}[LB]{lower-band}
\newacro{ub}[UB]{upper-band}
\newacro{bl}[BL]{band-limited}
\newacro{bb}[BB]{broadband}
\newacro{rbms}[RBMs]{restricted Boltzmann machines}
\newacro{gmm}[GMM]{Gaussian mixture model}
\newacro{lstm}[LSTM]{long short-term memory}
\newacro{blstm}[BLSTM]{bidirectional LSTM}
\newacro{lpcc}[LPCC]{linear predictive cepstral coefficient}
\newacro{bidilconv}[Bi-DilConv]{bidirectional dilated convolution}
\newacro{stfc}[STFC]{short-time Fourier convolution}
\newacro{bsft}[BSFT]{bandwidth spectral feature transform}
\newacro{udm}[UDM]{unconditional diffusion model}
\newacro{sgm}[SGM]{score-based generative model}
\newacroplural{sgm}[SGMs]{score-based generative models}

\newacro{ldm}[LDM]{latent diffusion model}
\newacro{kl}[KL]{Kullback-Leibler}
\newacro{hrnn}[HRNN]{hierarchical RNN}
\newacro{lsgan}[LSGAN]{least-squares GAN}
\newacro{nac}[NAC]{neural audio
codec}
\newacro{1d}[1-D]{one-dimensional}
\newacro{2d}[2-D]{two-dimensional}

%% file: sec/0_abstract.tex
\begin{abstract}
Audio super-resolution (SR), also referred to as bandwidth extension (BWE), aims to reconstruct high-fidelity signals from low-resolution (LR) or band-limited (BL) observations, an inherently ill-posed task due to the ambiguity of missing high-frequency (HF) content. This survey provides a comprehensive overview of the field, with a particular focus on the paradigm shift from discriminative mapping to modern generative modeling. We first review early discriminative deep neural network (DNN) models, which formulate BWE/SR as a deterministic mapping problem and are prone to regression-to-the-mean effects and spectral over-smoothing.
We then systematically review generative approaches, including autoregressive (AR) models, variational autoencoders (VAEs), generative adversarial networks (GANs), diffusion and score-based models, flow-based methods, and Schrödinger bridges. Across these approaches, we examine key design aspects, including representation domain, architecture, conditioning mechanisms, and trade-offs among reconstruction fidelity, perceptual quality, robustness, and computational efficiency. We further conduct unified experiments on representative discriminative and generative methods to provide controlled empirical evidence for these trade-offs.
Furthermore, we discuss emerging directions involving large language models (LLMs) and multimodal foundation models, and highlight open challenges in perceptual evaluation, practical deployment, and real-world generalization. By providing a structured taxonomy and unified perspective, this survey establishes a comprehensive foundation and offers a practical roadmap for advancing BWE/SR from deterministic point estimation toward distribution-aware generative modeling.
\end{abstract}

%% file: sec/1_introduction_final.tex
\section{Introduction}
\Ac{asr}~\citep{kuleshov2017audio} and \ac{bwe}~\citep{abel2017artificial} are essential tasks in artificial intelligence and audio research communities. They are widely used in applications such as telecommunications~\citep{li2015deep}, hearing aids~\citep{van2020speech}, and legacy recording restoration~\citep{moliner2024blind}, and can benefit downstream tasks including speech and speaker recognition~\citep{li2019speech,yamamoto2019speaker,yang2023time,yang2026towards}. Although these terms are defined from different perspectives and arise from different historical contexts, they share the common objective of restoring high-fidelity audio from bandwidth-constrained observations. From a learning perspective, \ac{bwe}/\ac{asr} can be formulated as ill-posed reconstruction problems in which missing \ac{hf} content must be inferred from  \ac{bl} or \ac{lr} observations. Because the missing \ac{hf} content is not uniquely determined by the input, a single degraded waveform may correspond to multiple perceptually plausible \ac{bb} or \ac{hr} reconstructions. This inherent one-to-many ambiguity makes \ac{bwe}/\ac{asr} a useful testbed for conditional distribution modeling beyond deterministic point estimation.

Traditional techniques, including source-filter models~\citep{makhoul1979high}, codebook mapping~\citep{unno2005robust}, \acp{gmm}~\citep{ohtani2014gmm}, and \acp{hmm}~\citep{chen2004hmm}, relied on structured priors and simplified acoustic assumptions, often producing over-smoothed spectra and audible artifacts due to limited modeling capacity.
The advent of \ac{dl}~\citep{lecun2015deep,liu2024m4x,jia2025end,li2026sepprune} shifted \ac{bwe}/\ac{asr} toward data-driven modeling of complex spectral relationships. Early \ac{dnn} architectures, such as \acp{mlp}~\citep{li2015dnn,liu2015novel,wang2015speech}, \acp{rnn}~\citep{schmidt2018blind,birnbaum2019temporal,hou2020speaker}, and \acp{cnn}~\citep{kuleshov2017audio,lim2018time,wang2020time,tamiti2025high,jia2026adaptive}, demonstrated superior performance over statistical models by directly learning non-linear mappings from degraded observations to clean targets. Subsequent research introduced more sophisticated structures, including U-Net variants~\citep{kuleshov2017audio}, Transformers~\citep{vaswani2017attention}, and Mambas~\citep{shams2024ssamba}, to better capture fine-grained spectral details and model long-range temporal dependencies. However, despite their success, these discriminative models frequently suffer from regression-to-the-mean behavior when trained with standard distance-based losses, resulting in spectral over-smoothing and a lack of \ac{hf} richness. These limitations are especially pronounced in \ac{bwe}/\ac{asr}, where the missing \ac{hf} content is inherently ambiguous and may admit multiple plausible reconstructions.

\begin{figure*}[!t]
\centering
\includegraphics[width=0.99\textwidth]{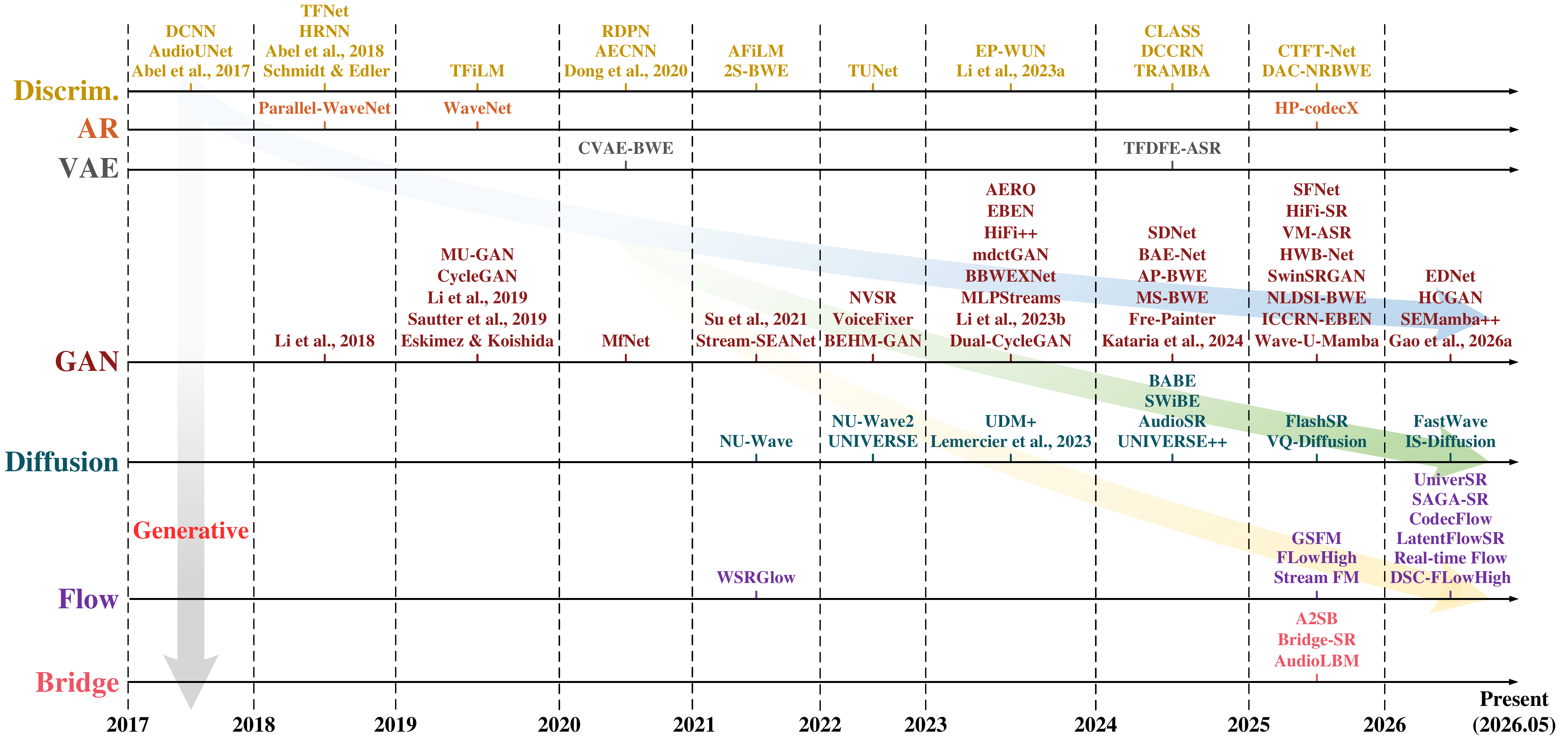}
\caption{\textbf{Timeline of methodological evolution in \ac{bwe} and \ac{asr} (2017–present).} The trajectory highlights a clear generative tendency: after the early dominance of deterministic models, modern likelihood-based or score-based generative approaches are increasingly shaping the state-of-the-art paradigm, reflecting a shift from point estimation to conditional distribution matching for perceptually plausible \ac{hf} reconstruction. 
}
\label{fig:timeline}
\end{figure*}

Recent advances in generative modeling have provided new paths to address this ambiguity by modeling the conditional distribution of plausible \ac{hf} content rather than predicting a single deterministic target.
Earlier generative models, including \ac{ar} models~\citep{gupta2019speech} and \acp{vae}~\citep{bachhav2020artificial}, support probabilistic generation through output or latent-variable sampling. Yet these methods face distinct trade-offs: \ac{ar} models provide fine-grained temporal modeling at the cost of high inference latency, while \acp{vae} offer structured latent spaces but often lack \ac{hf} fidelity. Later, \acp{gan}~\citep{li2018speech,eskimez2019speech} are adopted to enhance spectral sharpness through adversarial training, although many remain deterministic at inference and are prone to training instability and mode collapse.
More recently, research has increasingly explored likelihood- and score-based paradigms, such as diffusion probabilistic models~\citep{ho2020denoising,nichol2021improved,serra2022universal}, flow-based methods~\citep{lipman2023flowmatch,tong2023improving,gat2024discrete}, and Schrödinger bridges~\citep{de2021diffusion,shi2023diffusion,li2025bridge}, which provide alternative ways to model conditional audio distributions through iterative denoising, continuous-time dynamics, or stochastic bridges between degraded and clean signals. These methods have broadened the design space for \ac{bwe}/\ac{asr}, introducing new trade-offs among sample quality, diversity, controllability, computational cost, and inference latency. To contextualize this modeling shift from discriminative formulations to generative designs, Figure~\ref{fig:timeline} summarizes representative \ac{bwe}/\ac{asr} studies over recent years.

Despite the rapid advancements in \ac{bwe}/\ac{asr}, the field currently lacks a systematic work that articulates the significant paradigm shift from discriminative mapping to modern generative modeling frameworks. 
Existing tutorials and surveys~\citep{larsen2005audio, prasad2016bandwidth} remain fragmented and increasingly outdated: they primarily emphasize classical signal-processing approaches and provide limited coverage of \ac{dnn}-based models. Moreover, the definitional relationship between \ac{asr} and \ac{bwe} remains underexplored, leaving their distinctions and common objectives ambiguous across communities. To the best of our knowledge, this survey addresses this critical lack by providing the \textit{\textbf{first comprehensive overview}} that tracks the evolution of both \ac{bwe} and \ac{asr} through this lens. We systematically organize the major classes of generative models, highlighting the shift from simple point estimation to modeling complex, high-dimensional audio distributions. We conduct unified experiments across representative discriminative and generative methods to quantitatively compare reconstruction accuracy, perceptual quality, model complexity, and inference efficiency. By distilling the architectural principles, conditioning mechanisms, and inherent trade-offs of these approaches, this work also provides practical guidance for designing high-fidelity audio systems. Finally, we identify emerging directions, including \acp{llm} and multimodal foundation models~\citep{yang2025retrieval,xu2025synthetic,liu2025rl,yang2025m,xiong2026enhancing,xiong2026adaptive}, and highlight open challenges in perceptual evaluation, computational efficiency, robustness, and practical deployment. Our contributions are summarized below:
\begin{itemize}
    \item We formulate \ac{bwe} and \ac{asr} under a unified framework of \ac{hf} reconstruction and clarify the relationship and distinctions between the two terms.
    \item We organize the literature by modeling paradigm and architectural design, tracing the progression from discriminative neural networks to modern generative approaches.
    \item We conduct unified experiments on representative discriminative and generative methods to compare trade-offs in reconstruction accuracy, perceptual quality, model complexity, and inference efficiency.
    \item We discuss emerging trends and highlight open challenges to guide future research.
\end{itemize}

The remainder of this paper is organized as follows. Section~\ref{sec:problem} introduces the problem formulation and survey scope. Section~\ref{sec:learning} establishes the learning framework for \ac{bwe}/\ac{asr}, including input representations, training targets \& objective functions, and spectral mapping paradigms. Section~\ref{sec:eval} reviews benchmark datasets and evaluation metrics. Section~\ref{sec:dis_model} surveys discriminative modeling approaches, while Section~\ref{sec:gen model} provides an in-depth analysis of generative methods. Section~\ref{sec:quan} presents a unified quantitative comparison of representative model families, and Section~\ref{sec:challenge} discusses the remaining challenges, future directions, and broader impacts. Section~\ref{sec:literature_selection} describes the literature selection process. Finally, Section~\ref{sec:con} concludes this survey paper.

%% file: sec/2_background_final.tex
\begin{figure*}[t]
\centering
\includegraphics[width=0.995\textwidth]{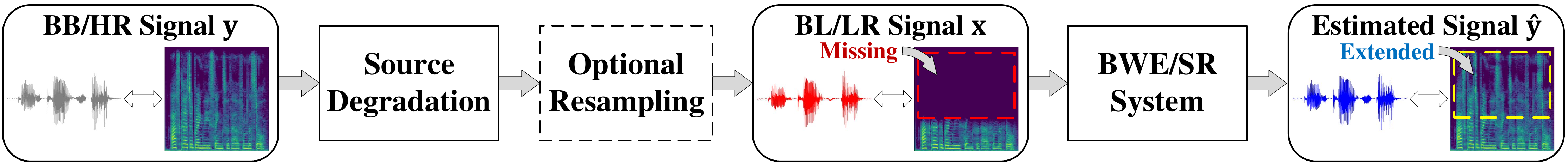}
\caption{\textbf{Signal flow diagram of \ac{bwe}/\ac{asr}.} The degradation process removes \ac{hf} spectral bandwidth from a reference signal $\mathbf{y}$, followed by an optional resampling stage, to produce the observation $\mathbf{x}$. In practical applications, this reference signal is not available. The \ac{bwe}/\ac{asr} system then estimates the reconstruction $\hat{\mathbf{y}}$ from the observation $\mathbf{x}$. Waveforms and spectrograms at each stage visualize the transition from the reference signal to the degraded input and finally to the recovered signal.}
\label{fig:signalflow}
\end{figure*}

\section{Background}
\label{sec:problem}

\subsection{Problem Formulation}
\label{subsec:problem}

The tasks of \ac{bwe}, \ac{asr}, and \ac{ssr} can each be formulated as learning a function
\begin{equation}
\label{eq:bwe_mapping}
f:\; \mathbf{x} \mapsto \mathbf{y},
\end{equation}
where $\mathbf{x} = [x[1], \dots, x[N_1]]^\top $ and $\mathbf{y} = [y[1], \dots, y[N_2]]^\top$ are \ac{1d} discrete audio sequences with $N_1$ and $N_2$ samples, respectively, and $x[n]$ and $y[n]$ denote the corresponding scalar-valued samples of $\mathbf{x}$ and $\mathbf{y}$.
For \ac{bwe}, the core objective is to restore missing \ac{hf} spectral components, where the input $\mathbf{x}$ and target $\mathbf{y}$ correspond to \ac{bl} and \ac{bb} signals, respectively.
In simulation, $\mathbf{x}$ is typically obtained by applying a low-pass filter to $\mathbf{y}$ to remove spectral content above a specific cutoff frequency.
Some studies~\citep{wang2015speech,liu2016novel,abel2018simple,li2021real,hauret2023eben,li2023restoration} perform \ac{bwe} by maintaining identical sampling rates for inputs and targets, resulting in equal sequence lengths ($N_1 = N_2$). In this configuration, the \ac{bl} signal is represented on the same temporal grid as the target \ac{bb} signal. Conversely, others~\citep{liu2015novel,li2015deep,gu2016speech, dong2020time, bachhav2020artificial,su2021bandwidth} use a lower sampling rate for the input, resulting in $N_1 < N_2$. This requires the model to resolve temporal resolution mismatches in addition to spectral recovery, thereby aligning the formulation closely with \ac{asr}.
Conventional \ac{bwe} assumes a known, fixed degradation process (e.g., a predefined band-limiting filter), whereas blind \ac{bwe}~\citep{moliner2024blind} relaxes this assumption and must infer unknown degradations from the observed signal.
Although \ac{bwe} has historically focused on speech signals to improve intelligibility in telecommunications, recent advancements have extended these techniques to musical applications~\citep{moliner2022behm,moliner2024blind,hernandez2026single}. 

Conversely, the term \ac{asr} emerged more recently as a direct conceptual parallel to image super-resolution~\citep{dong2016image,yue2016image,yang2019deep} in computer vision. In the \ac{asr} setting, the input $\mathbf{x}$ is a \ac{lr} signal sampled at rate $R$, and the target $\mathbf{y}$ is the corresponding \ac{hr} signal sampled at a higher rate $R' > R$, with the upsampling ratio defined as
\begin{equation}
\label{eq:upsampling_ratio}
r = \frac{R'}{R}.
\end{equation}
Consequently, \ac{asr} is explicitly framed as a temporal interpolation problem where the model must increase the sampling rate, mapping the input sequence of length $N_1$ to an output of length $N_2 = rN_1$.
Under the assumption that $\mathbf{x}$ and $\mathbf{y}$ are uniformly sampled from finite-bandwidth continuous-time signals, the Nyquist-Shannon sampling theorem~\citep{jerri1977shannon} guarantees that the maximum representable bandwidth is $R/2$ and $R'/2$, respectively. Therefore, \ac{asr} inherently requires the restoration of \ac{hf} content, in contrast to a standard upsampling operation that only increases the sampling rate. Without additional \ac{hf} recovery, simply performing \ac{bl} interpolation would provide no new information above the original Nyquist limit $R/2$. Additionally, \ac{ssr} is defined similarly to \ac{asr} but is restricted to speech signals, whereas \ac{asr} targets general audio, including speech, music, and sound effects.

Fig.~\ref{fig:signalflow} illustrates the signal flow for the degradation and reconstruction processes of \ac{bwe} and \ac{asr}, where both waveforms and spectrograms are shown at each stage to visualize the signal transformations. Table~\ref{tab:problem_definitions} summarizes the key differences and commonalities between \ac{bwe}, \ac{asr}, and \ac{ssr}. While \ac{bwe} is defined by the restoration of spectral bandwidth, \ac{asr} and \ac{ssr} are characterized by temporal interpolation. Despite these differing definitions, all three tasks can be unified as ill-posed inverse problems—specifically spectral or temporal inpainting. In this framework, missing \ac{hf} information is statistically reconstructed through data-driven modeling to approximate the natural audio characteristics learned from training data. Throughout this paper, we use \ac{bwe}, \ac{asr}, and \ac{ssr} to refer to specific works, while the notation \ac{bwe}/\ac{asr} collectively denotes the broader research field.

\begin{table}[t]
\centering
\caption{\textbf{Comparison of \ac{bwe}, \ac{asr}, and \ac{ssr}} in terms of their underlying degradation processes, input-target relationships, sampling rate changes, primary reconstruction objectives, and application domains.}
\vspace{0pt}
\label{tab:problem_definitions}
\renewcommand{\arraystretch}{1.5}
\small
\begin{tabularx}{\textwidth}{c c c c c c}
\toprule
\textbf{Task} &
\textbf{Source Degradation} &
\textbf{Input $\rightarrow$ Target} &
\textbf{Rate} &
\textbf{Primary Objective} &
\textbf{Signal Scope} \\
\midrule
\ac{bwe} &
Bandwidth Reduction &
\ac{bl} $\rightarrow$ \ac{bb} Signal &
$\uparrow$ / = &
Restoring Bandwidth &
Primarily Speech \\

\ac{asr} &
Temporal Downsampling &
\ac{lr} $\rightarrow$ \ac{hr} Signal &
$\uparrow$ &
Restoring Time Resolution &
General Audio \\

\ac{ssr} &
Temporal Downsampling &
\ac{lr} $\rightarrow$ \ac{hr} Signal &
$\uparrow$ &
Restoring Time Resolution &
Speech Only \\
\bottomrule
\end{tabularx}
\end{table}

\subsection{Survey Scope}
\label{subsec:survey_scope}
To maintain a focused survey of \ac{bwe} and \ac{asr}, we establish specific scope constraints regarding the input signal and restoration task. When the input $\mathbf{x}$ is both \ac{bl} and corrupted by additive noise, the task is defined as noise-robust \ac{bwe}~\citep{lin2023noise,yang2024sdnet,lin2024swibe, liu2025neural,liu2026latentflowsr}. 
Furthermore, universal \ac{se}, also referred to as speech restoration, has recently emerged as a prominent research direction, as promoted by the URGENT challenge~\citep{zhang2024urgent}. Such approaches~\citep{serra2022universal,liu2022voicefixer,andreev2023hifiplusplus,lemercier2023analysing,scheibler2024universal,ku2025generative,welker2025real,kwak2026ednet,gao2026real,lee2026semambaplusplus,hsieh2026towards} are designed to handle diverse distortions, including additive noise, reverberation, clipping, and bandwidth limitation. We include these works as they remain within the single-observation inverse problem framework.
Conversely, we exclude methods that leverage multiple input modalities, such as the fusion of bone-conducted and air-conducted speech~\citep{yu2020time,wang2022fusing}. We categorize these approaches as multi-modal \ac{se} rather than pure \ac{bwe}/\ac{asr}, as their performance relies on sensor fusion rather than the data-driven spectral modeling considered in this work.

%% file: sec/3_training.tex
\section{Learning Framework}
\label{sec:learning}
\subsection{Input Representations}
\label{subsec:feature}
Early \ac{dnn}-based methods have primarily relied on spectral-domain features that explicitly characterize the short-time spectral shape and excitation information of speech signals.
Typical examples include the \ac{lps}~\citep{li2015deep,li2015dnn,liu2015novel,gu2016speech,liu2016novel,abel2018simple,schmidt2018blind,eskimez2019speech,bachhav2020artificial}, 
\ac{lsf}~\citep{liu2015novel,liu2016novel,li2018speech}, 
\ac{mfcc}~\citep{liu2015novel,liu2016novel},
and band-pass voicing coefficient (BPVC)~\citep{liu2015novel,liu2016novel}.
Beyond these frequency-domain descriptors, other works~\citep{abel2016artificial,abel2017artificial} explore time-domain and modulation-domain features such as the autocorrelation function, zero-crossing rate, gradient index, and normalized relative frame energy. In addition to these handcrafted features, some approaches instead learn compact spectral-envelope representations directly from data, using methods such as the Gaussian-Bernoulli restricted Boltzmann machine~\citep{wang2015speech}, STRAIGHT vocoder~\citep{gu2015restoring}, and a specially structured \ac{dnn}~\citep{gu2016speech}.

The advent of \ac{dnn} architectures has motivated a shift toward learning directly from time-domain and frequency-domain audio representations. In the literature, numerous works operate directly on the raw waveform \citep{kuleshov2017audio,gu2017waveform,lim2018time,ling2018waveform,birnbaum2019temporal,dong2020time,hou2020speaker,wang2020time,wang2021towards,rakotonirina2021self,li2021real,nguyen2022tunet,kim2024class}, leveraging convolutional encoders to learn rich temporal-spectral representations while implicitly preserving phase information and avoiding the constraints and preprocessing artifacts associated with fixed time-frequency transforms. A special case is EBEN \citep{hauret2023eben}, which operates on subband waveforms derived from \ac{pqmf} decomposition. In addition to taking waveforms as the input, \citep{lim2018time} and \citep{dong2020time} further incorporate the \ac{stft} magnitude spectrogram and enable dual-branch modeling in the time and frequency domains.
Additionally, log-mel spectrograms and mel-spectrograms~\citep{gupta2019speech,liu2022neural,liu2022voicefixer,andreev2023hifiplusplus,kim2024audio,liu2024audiosr,im2025flashsr,zhao2025hifi,yun2025flowhigh} have been widely used due to their strong alignment with human auditory perception. They are computed by mapping \ac{stft} magnitudes through perceptually spaced mel filterbanks, with or without an additional logarithmic transformation, respectively. However, both \ac{stft} magnitude spectrograms and mel-spectrograms lack the phase information and therefore require additional algorithms for phase recovery.

To address this limitation, several studies instead use \ac{stft} complex spectrograms~\citep{mandel2023aero,soltanmohammadi2023low,edraki2024speaker,lu2024towards,lu2024multi,yu2024bae,tamiti2025high,ku2025generative,zhang2025vm}, capturing both real and imaginary components—or equivalently, log-magnitude and phase spectra—to enable more accurate waveform reconstruction and phase-aware modeling. Moreover, recent studies~\citep{liu2021swin,shuai2023mdctgan} employ the \ac{mdct} and leverage the real-valued spectrogram representation. This technique supports phase-aware reconstruction by encoding phase information into the sign of the \ac{mdct} coefficients. \ac{mdct} removes inherent redundancy in the complex spectrogram and operates with fewer coefficients for the same temporal resolution, providing a compact and computationally efficient solution. Finally, recent trends~\citep{fang2025vector,liu2025neural,ginies2025harmonic,zhang2026codecflow} employ discrete latent tokens extracted from a \ac{nac} as features, producing a compact and semantically rich representation that enables high-fidelity and perceptually consistent audio reconstruction. 
Overall, the evolution of input representations reflects a trajectory from handcrafted spectral features toward richer, phase-aware, and semantically grounded representations, driven by both architectural advances and increasing demands of high-fidelity reconstruction.


\subsection{Training Targets and Objective Functions}
\label{subsec:target}
Early \ac{bwe} methods have relied on spectral-domain regression, which predicts the \ac{hf} or fullband \ac{lps}~\citep{li2015deep,li2015dnn, liu2015novel,gu2016speech, liu2016novel,eskimez2019speech}, \ac{hf} cepstral vector~\citep{abel2017artificial,abel2018simple}, \ac{hf} posterior probabilities modeled by \ac{hmm}~\citep{abel2016artificial}, or other spectral-envelope representations~\citep{wang2015speech,gu2015restoring,schmidt2018blind,li2018speech, bachhav2020artificial}. These approaches are typically trained using the \ac{mmse} criterion. The missing phase information is reconstructed by mirroring the \ac{lf} phase, followed by \ac{istft} and overlap-add synthesis. Such spectral-domain training targets provide a smooth and meaningful representation that is easier to model and leads to stable training. However, they discard phase information and cannot capture the stochastic structure of \ac{hf} components, often resulting in over-smoothed spectra and limited perceptual realism.

Given these limitations, \ac{bwe}/\ac{asr} research has shifted toward direct waveform prediction~\citep{kuleshov2017audio,gu2017waveform,wang2018speech,lim2018time,ling2018waveform,birnbaum2019temporal,gupta2019speech,li2019speech,wang2020time,hou2020speaker,dong2020time,hao2020time,wang2021towards,li2021real,su2021bandwidth,rakotonirina2021self,nguyen2022tunet,li2023restoration,andreev2023hifiplusplus,kim2024class,zhao2025hifi,lee2025wave}, where the model implicitly learns both amplitude and phase and can capture fine-grained \ac{hf} details without relying on explicit spectral synthesis. Typical time-domain loss functions include the \ac{mse} loss~\citep{kuleshov2017audio,lim2018time,birnbaum2019temporal,hao2020time,rakotonirina2021self,kim2024class} and \ac{mae} loss~\citep{li2019speech,wang2020time,dong2020time,wang2021towards,su2021bandwidth,li2023restoration} applied directly to the waveform, \ac{sisdr} loss~\citep{hou2020speaker}, and \ac{ce} loss~\citep{gu2017waveform,wang2018speech,ling2018waveform,gupta2019speech}. In addition, several studies incorporate frequency-domain losses for waveform targets, including \ac{mse} loss on \ac{mfcc}~\citep{li2019speech}, \ac{mae} loss on \ac{stft} spectrograms~\citep{wang2020time,wang2021towards,su2021bandwidth}, \ac{mae} loss on mel or log-mel spectrograms~\citep{hao2020time,su2021bandwidth,zhao2025hifi,lee2025wave}, multi-scale spectral energy loss~\citep{li2023restoration}, and multi-scale \ac{stft} loss~\citep{nguyen2022tunet,kim2024class,lee2025wave}. Later research has shown that combining time-domain and frequency-domain losses provides complementary constraints that enforce both sample-level waveform accuracy and perceptually important spectral structure. 

Parallel to waveform prediction, another major class of approaches uses the \ac{stft} complex spectrograms~\citep{mandel2023aero,soltanmohammadi2023low,yang2024sdnet,edraki2024speaker,lu2024towards,lu2024multi,yu2024bae,tamiti2025high,zhang2025vm} or \ac{mdct} spectrograms~\citep{shuai2023mdctgan,yuan2025swinsrgan} as training targets, allowing models to explicitly or implicitly account for phase information in the spectral domain.
Similar loss functions are used, including the \ac{mae} loss~\citep{soltanmohammadi2023low,edraki2024speaker,yu2024bae} and \ac{sisdr} loss~\citep{tamiti2025high} applied to the waveform, while the most widely adopted frequency-domain objective is the multi-scale \ac{stft} loss~\citep{mandel2023aero,soltanmohammadi2023low,yang2024sdnet,yu2024bae,tamiti2025high,zhang2025vm}. Moreover, AP-BWE~\citep{lu2024towards} and its subsequent extension~\citep{lu2024multi} introduce a combination of log-amplitude \ac{mse} loss, phase anti-wrapping loss, and short-time complex spectral \ac{mse} loss to jointly model amplitude and phase. Additionally, several approaches~\citep{liu2022neural,liu2022voicefixer,liu2024audiosr,kim2024audio,im2025flashsr,yun2025flowhigh} adopt a two-stage pipeline where the model first predicts a mel-spectrogram as an intermediate representation and then applies a pretrained vocoder to synthesize the waveform. In the first stage, the \ac{mae} loss is typically used to achieve accurate mel-spectrogram prediction.

\begin{figure*}[t]
\label{fig:bwe_taxonomy}
\centering
\resizebox{\textwidth}{!}{%
\begin{forest}
for tree={
grow=east,
reversed=true,
forked edge,
anchor=base west,
parent anchor=east,
child anchor=west,
base=left,
font=\large,
rectangle,
draw=black,
rounded corners,
align=center,
minimum width=4em,
edge={black,line width=0.8pt},
s sep=4pt,
l sep=8pt,
inner xsep=2pt,
inner ysep=2.0pt,
line width=0.8pt,
ver/.style={rotate=90,child anchor=north,parent anchor=south,anchor=center}
},
where level=1{text width=6.0em,font=\tiny}{},
where level=2{text width=8.7em,font=\tiny}{}
[\textbf{BWE/SR} \\\textbf{Systems}, rootnode,
    [\textbf{16 kHz},rate node
        [\textbf{Fixed-constraint},fixed node
            [{\litbox{
            \citet{li2015dnn},
            \citet{li2015deep},
            \citet{wang2015speech},
            \citet{liu2015novel},
            \citet{gu2015restoring},
            \citet{abel2016artificial},
            \citet{liu2016novel},
            \citet{gu2016speech},
            \citet{abel2017artificial},
            DCNN~\citep{gu2017waveform},
            ~\citet{abel2018simple},
            ~\citet{li2018speech},
            HRNN~\citep{ling2018waveform},
            Parallel-WaveNet~\citep{wang2018speech},
            ~\citep{li2019speech},
            CycleGAN~\citep{haws2019cyclegan},
            ~\citep{dong2020time},
            CVAE-BWE~\citep{bachhav2020artificial},
            AECNN~\citep{wang2020time,wang2021towards},
            MfNet~\citep{hao2020time},
            Stream-SEANet~\citep{li2021real},
            2S-BWE~\citep{lin2021two},
            TUNet~\citep{nguyen2022tunet}, MLPStreams~\citep{soltanmohammadi2023low},
            mdctGAN~\citep{shuai2023mdctgan},
            UDM+~\citep{yu2023conditioning},
            ~\citet{li2023two},
            ~\citet{li2023restoration},
            SWiBE~\citep{lin2024swibe},
            CLASS~\citep{kim2024class},
            DCCRN~\citep{edraki2024speaker},
            TRAMBA~\citep{sui2024tramba},
            DAC-NRBWE~\citep{liu2025neural},
            ~\citet{kim2025progressive}, CodecFlow~\citep{zhang2026codecflow}
            }}, fixed leaf]
        ]
        [\textbf{Multi-scenario},multi node
            [{\litbox{
            AudioUNet~\citep{kuleshov2017audio},
            TFNet~\citep{lim2018time},
            ~\citet{eskimez2019speech},
            TFiLM~\citep{birnbaum2019temporal},
            MU-GAN~\citep{kim2019bandwidth},
            RDPN~\citep{hou2020speaker},
            AFiLM~\citep{rakotonirina2021self},
            NVSR~\citep{liu2022neural},
            HiFi++~\citep{andreev2023hifiplusplus},
            ~\citet{lemercier2023analysing},
            AERO~\citep{mandel2023aero},
            TFDFE-ASR~\citep{tian2024time},
            EDNet~\citep{kwak2026ednet},
            HCGAN~\citep{gao2026harmonic}
            }}, multi leaf]
        ]
        [\textbf{Bandwidth-agnostic},agno node
            [{\litbox{
            UNIVERSE~\citep{serra2022universal}, EBEN~\citep{hauret2023eben},
            SDNet~\citep{yang2024sdnet},
            AP-BWE~\citep{lu2024towards},
            GSFM~\citep{ku2025generative},
            VM-ASR~\citep{zhang2025vm},
            NLDSI-BWE~\citep{tamiti2025nldsi},
            ICCRN-EBEN~\citep{bao2025frequency}, SEMamba++~\citep{lee2026semambaplusplus}
            }},agno leaf]
        ]
    ]
    [\textbf{22.05 kHz},rate node
        [\textbf{Bandwidth-agnostic},agno node
            [{\litbox{
            BEHM-GAN~\citep{moliner2022behm},
            BABE~\citep{moliner2024blind},
            ~\citep{gao2026real}
            }},agno leaf]
        ]
    ]
    [\textbf{24 kHz},rate node
        [\textbf{Fixed-constraint},fixed node
            [{\litbox{
            WaveNet~\citep{gupta2019speech},
            AERO~\citep{mandel2023aero},
            UNIVERSE++~\citep{scheibler2024universal},
            TFDFE-ASR~\citep{tian2024time}
            }}, fixed leaf]
        ]
        [\textbf{Multi-scenario},multi node
            [{\litbox{
            Fre-Painter~\citep{kim2024audio},
            IS-Diffusion~\citep{jin2026inference}
            }},multi leaf]
        ]
    ]
    [\textbf{32 kHz},rate node
        [\textbf{Fixed-constraint},fixed node
            [{\litbox{
            BBWEXNet~\citep{gomez2023low}
            }},fixed leaf]
        ]
    ]
    [\textbf{44.1 kHz},rate node
        [\textbf{Fixed-constraint},fixed node
            [{\litbox{
            AERO~\citep{mandel2023aero},
            TFDFE-ASR~\citep{tian2024time},
            CodecFlow~\citep{zhang2026codecflow}
            }},fixed leaf]
        ]
        [\textbf{Bandwidth-agnostic},agno node
            [{\litbox{
            VoiceFixer~\citep{liu2022voicefixer},
            NVSR~\citep{liu2022neural},
            A2SB~\citep{shihaudio},
            SAGA-SR~\citep{im2026saga},
            DSC-FLowHigh~\citep{hernandez2026single},
            LatentFlowSR~\citep{liu2026latentflowsr}
            }},agno leaf]
        ]
    ]
    [\textbf{48 kHz},rate node
        [\textbf{Fixed-constraint},fixed node
            [{\litbox{
            AERO~\citep{mandel2023aero},
            BBWEXNet~\citep{gomez2023low},
            Dual-CycleGAN~\citep{yoneyama2023nonparallel}
            }},fixed leaf]
        ]
        [\textbf{Multi-scenario},multi node
            [{\litbox{
            \citet{su2021bandwidth},
            NU-Wave~\citep{lee2021nu},
            WSRGlow~\citep{zhang2021wsrglow},
            mdctGAN~\citep{shuai2023mdctgan},
            UDM+~\citep{yu2023conditioning},
            Fre-Painter~\citep{kim2024audio},
            MS-BWE~\citep{lu2024multi},
            BAE-Net~\citep{yu2024bae},
            TFDFE-ASR~\citep{tian2024time},
            CTFT-Net~\citep{tamiti2025high},
            SFNet~\citep{dai2025sfnet}
            }},multi leaf]
        ]
        [\textbf{Bandwidth-agnostic},agno node
            [{\litbox{
            NU-Wave2~\citep{han2022nu},
            AudioSR~\citep{liu2024audiosr},
            AP-BWE~\citep{lu2024towards},
            AudioLBM~\citep{li2025audio},
            Bridge-SR~\citep{li2025bridge},
            FlashSR~\citep{im2025flashsr},
            HiFi-SR~\citep{zhao2025hifi},
            FLowHigh~\citep{yun2025flowhigh},
            ~\citep{tamiti2025high},
            VM-ASR~\citep{zhang2025vm},
            VQ-Diffusion~\citep{fang2025vector},
            Wave-U-Mamba~\citep{lee2025wave},
            SwinSRGAN~\citep{yuan2025swinsrgan},
            NLDSI-BWE~\citep{tamiti2025nldsi},
            SFNet~\citep{dai2025sfnet},
            UniverSR~\citep{choi2025universr},
            HWB-Net~\citep{liu2025hwb},
            FastWave~\citep{kuznetsov2026fastwave}
            }},agno leaf]
        ]
    ]
    [\textbf{96 or 192 kHz},rate node
        [\textbf{Bandwidth-agnostic},agno node
            [{\litbox{
            AudioLBM~\citep{li2025audio}
            }},agno leaf]
        ]
    ]
]
\end{forest}%
}
\vspace{-0mm}
\caption{\textbf{Taxonomy of \ac{bwe}/\ac{asr} Literature.} Existing methods are organized by target sampling rates $\{16, 22.05, 24, 44.1, 48, 96, 192\}$ kHz and further categorized according to their spectral mapping paradigm \{fixed-constraint, multi-scenario, bandwidth-agnostic\} in training settings.}
\label{fig:final_bwe_taxonomy}
\vspace{-1em}
\end{figure*}
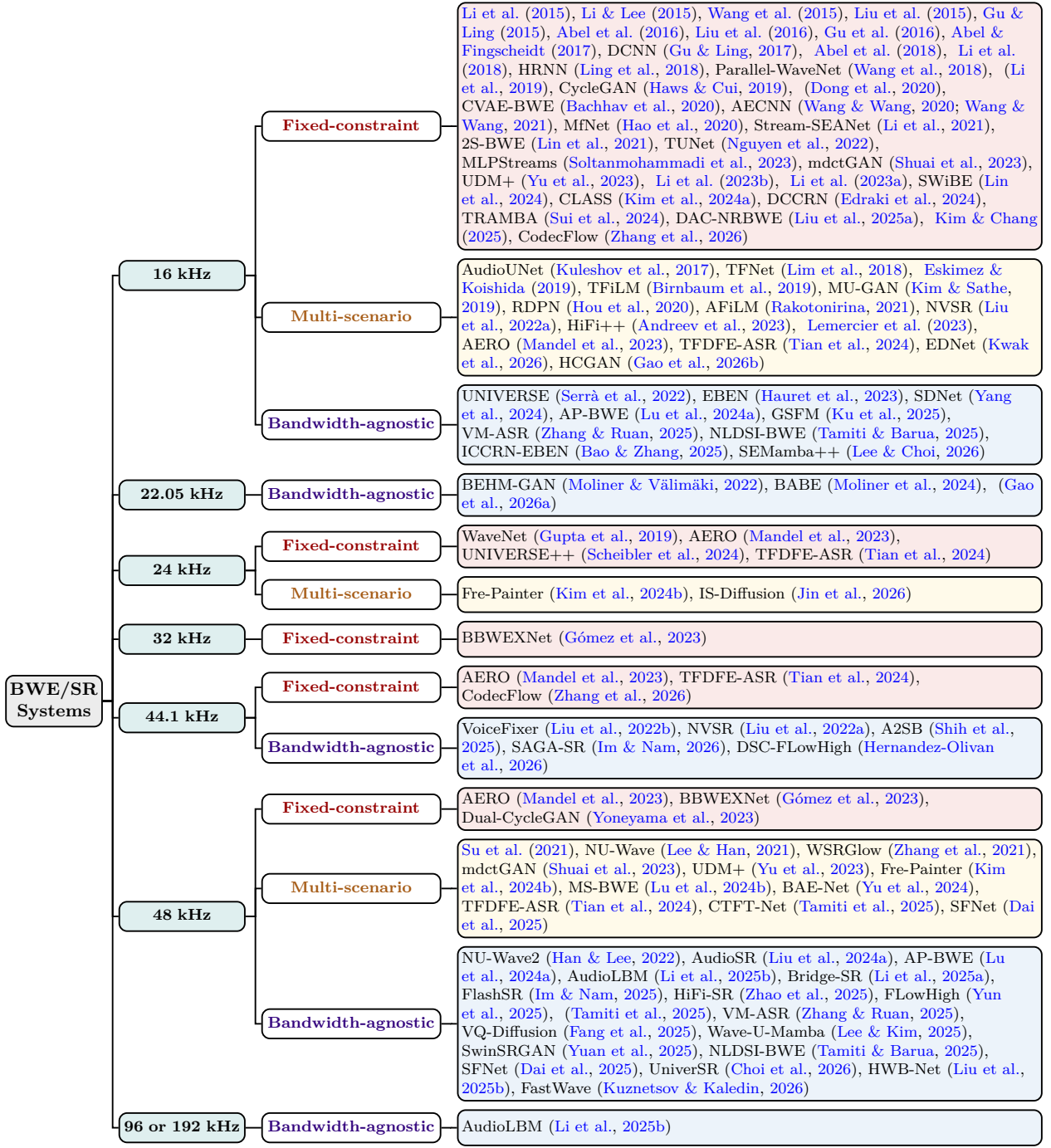

Regarding diffusion and score-based models~\citep{lee2021nu,han2022nu,yu2023conditioning,lin2024swibe,liu2024audiosr,im2025flashsr,fang2025vector}, the training targets are typically the additive noise or an equivalent denoising objective at each diffusion step. In this case, the model predicts either the noise or velocity vector in the waveform domain, spectrogram domain, or a learned latent space. These models are optimized using \ac{mae} or \ac{mse} losses applied in the corresponding domain. For \ac{gan}-based approaches, the training target can be either waveforms or spectrograms, and a combination of adversarial loss and reconstruction loss is employed to enhance perceptual quality. In flow-based approaches~\citep{yun2025flowhigh,ku2025generative}, training is performed by regressing an analytically defined velocity field that transports a simple base distribution to the target \ac{bb}/\ac{hr} signal distribution using the \ac{mse} loss. 
In bridge models~\citep{li2025bridge,li2025audio}, training aims to estimate a drift field that bridges a source and a target distribution by minimizing the \ac{kl} divergence between the induced path measure and a reference stochastic process under boundary constraints, either directly in signal space or within a learned latent representation space. Finally, other methods use alternative training targets such as \ac{pqmf}-decomposed subband signals~\citep{hauret2023eben} or latent token sequences~\citep{fang2025vector,liu2025neural,ginies2025harmonic,zhang2026codecflow}.
Taken together, the evolution of training targets mirrors the shift in model paradigms: early spectral-regression objectives gave way to direct waveform prediction, which in turn has been complemented by phase-aware spectral targets and the objective functions of modern generative frameworks.

\subsection{Spectral Mapping Paradigms}
\label{subsec:mapping}

Existing \ac{bwe}/\ac{asr} frameworks can be categorized by their spectral mapping paradigm, which describes how the input cutoff frequency is configured during training—whether fixed at a single value, varied across a discrete set, or treated as a continuous free parameter. Three paradigms can be distinguished: \textit{fixed-constraint}, \textit{multi-scenario}, and \textit{bandwidth-agnostic}. The fixed-constraint paradigm restricts the experimental protocol to a unique spectral mapping, in which a single cutoff frequency is used for both training and evaluation. This setting is typical of specialized systems such as legacy telephony restoration, where the speech bandwidth is extended from \ac{nb} (e.g., $300$~Hz–$3.4$~kHz or $0$–$4$~kHz) to \ac{wb} (e.g., $0$–$8$~kHz). The multi-scenario paradigm trains and evaluates separate models for multiple discrete input cutoff frequencies, enabling improved robustness across heterogeneous bandwidth conditions compared to a single fixed-cutoff setting. For instance, AudioUNet~\citep{kuleshov2017audio} employs distinct models for upscaling ratios of $2\times, 4\times,$ and $6\times$ at a target sampling rate of $16$ kHz, each optimized specifically for its corresponding input bandwidth. The bandwidth-agnostic paradigm employs a single universal model trained to handle a wide range of input cutoff frequencies, enabling many-to-one spectral bandwidth mapping within a unified framework. In this regime, input cutoff frequencies are either sampled from a continuous stochastic range or selected from a discrete set, as in NVSR~\citep{liu2022neural}, which samples cutoff frequencies over 1--16~kHz, and AP-BWE~\citep{lu2024towards}, which uses a discrete set $\{4, 6, 8, 12\}$~kHz. Recent works have favored this paradigm for its architectural robustness: the model implicitly infers the input spectral support and reconstructs the missing \ac{hf} components across arbitrary frequency gaps without prior knowledge of the degradation parameters.
Fig.~\ref{fig:final_bwe_taxonomy} illustrates a hierarchical taxonomy of \ac{bwe}/\ac{asr} methods based on target sampling rates and spectral mapping paradigms.

%% file: sec/4_evaluation.tex
\section{Evaluations}
\label{sec:eval}

\subsection{Audio Datasets}
\label{subsec:data}
Most existing \ac{bwe}/\ac{asr} approaches are primarily designed for speech signals and are commonly evaluated using standard speech corpora. In particular, the TIMIT Acoustic-Phonetic Continuous Speech Corpus~\citep{garofolo1988getting} and the CSTR Voice Cloning Toolkit (VCTK)~\citep{yamagishi2012english} are the most frequently adopted benchmarks due to their high-quality recordings and phonetic diversity.
The TIMIT database contains approximately $5.4$ hours of English speech from $630$ speakers recorded at a sampling rate of $16$ kHz with detailed phonetic and word-level annotations.
The VCTK corpus consists of roughly $44$ hours of English speech from $109$ speakers with diverse accents recorded at a sampling rate of $48$ kHz. Other studies employ a wider range of speech datasets, including LibriSpeech~\citep{panayotov2015librispeech}, Wall Street Journal (WSJ0) corpus~\citep{paul-baker-1992-design}, LibriTTS~\citep{zen2019libritts}, HiFi-TTS~\citep{bakhturina2021hi}, VoiceBank-DEMAND~\citep{botinhao2016investigating}, and LJSpeech~\citep{ito2017ljspeech}. The \ac{bl}/\ac{lr} signals are typically synthesized from these ground-truth \ac{bb}/\ac{hr} sources by applying low-pass filtering or downsampling to simulate bandwidth-limited acquisition.

Among \ac{ssr}-focused works, VCTK at $48$ kHz has emerged as the de facto benchmark, used both for training and evaluation by systems such as NU-Wave~\citep{lee2021nu}, AudioSR~\citep{liu2024audiosr}, and FlashSR~\citep{im2025flashsr}, making it the most consistently adopted corpus for \ac{ssr} despite its original design as a multi-speaker text-to-speech resource.
Conversely, several bone-conducted \ac{bwe} approaches incorporate real-world sensor data. \citet{edraki2024speaker} recorded a proprietary corpus of own-voice signals using paired air and in-ear microphones, while \citet{li2023restoration,li2023two} use the Elevoc Simultaneously-recorded Microphone/Bone-sensor (ESMB) speech corpus\footnote{\url{https://github.com/elevoctech/ESMB-corpus}}.

Beyond speech, \ac{bwe}/\ac{asr} has expanded to music and general sounds, requiring datasets that capture greater spectral complexity. Several studies~\citep{kuleshov2017audio,lim2018time,rakotonirina2021self,mandel2023aero,liu2024audiosr,moliner2024blind,im2025flashsr,li2025audio} incorporate diverse music datasets to evaluate performance on signals with richer spectral density and polyphonic structure than speech. Representative datasets include MUSDB18-HQ~\citep{MUSDB18HQ}, MoisesDB~\citep{pereira2023moisesdb}, MedleyDB~\citep{bittner2014medleydb}, SDS~\citep{manco2023song}, MAESTRO~\citep{hawthorne2018enabling}, COCOChorales~\citep{wu2022chamber}, and FMA-small~\citep{defferrard2016fma}. These datasets span a wide range of musical content, from solo recordings to multi-instrument mixtures, enabling rigorous evaluation of \ac{hf} reconstruction and polyphonic detail.
A subset of studies~\citep{tian2024time,liu2024audiosr,im2025flashsr,li2025audio} further considers datasets of general sounds, such as ESC-50~\citep{piczak2015esc}, FreeSound~\citep{mei2024wavcaps}, and ShipsEar~\citep{santos2016shipsear}. These datasets cover diverse non-stationary and stochastic acoustic conditions—from urban noise to underwater signals—often lacking the harmonic regularity found in speech or music.

\subsection{Evaluation Methods}
\label{subsec:metric}
\subsubsection{Objective Metrics}
\textbf{Signal-to-Noise Ratio (SNR):} Given $N$ samples of a reference signal $\mathbf{s} = [s[1], \dots, s[N]]^\top$
and its estimate $\hat{\mathbf{s}} = [\hat{s}[1], \dots, \hat{s}[N]]^\top$, SNR~\citep{box1988signal} is defined as
\begin{equation}
\text{SNR}(\mathbf{s}, \hat{\mathbf{s}}) = 10 \log_{10}
\left(
\frac{\lVert \mathbf{s} \rVert^2}{\lVert \mathbf{s} - \hat{\mathbf{s}} \rVert^2}
\right).
\label{eq:snr}
\end{equation}
SNR is a standard metric in the signal processing literature, commonly used to evaluate waveform distortion in the time domain.
To better capture frame-level variations, the segmental signal-to-noise ratio (SegSNR)~\citep{paliwal1987speech} computes the average SNR over short-time segments of an audio signal.
Let us denote the $l$-th frame of $\mathbf{s}$ as $\mathbf{s}_l$ and the $l$-th frame of
$\hat{\mathbf{s}}$ as $\hat{\mathbf{s}}_l$, $l = 1, \dots, L$, where $L$
represents the total number of frames. SegSNR is defined as
\begin{equation}
\text{SegSNR}(\mathbf{s}, \hat{\mathbf{s}}) = \frac{1}{L} \sum_{l=1}^{L}
10 \cdot \log_{10}\!\left(
\frac{\| \mathbf{s}_l \|^2}{\| \mathbf{s}_l - \hat{\mathbf{s}}_l \|^2}
\right).
\label{eq:segsnr}
\end{equation}
Both metrics are widely used but have limited sensitivity to perceptual quality in \ac{bwe}/\ac{asr}: because the generated \ac{hf} components are stochastic and may not align sample-by-sample with the reference even when perceptually convincing, high SNR does not guarantee good \ac{hf} reconstruction quality.

\textbf{Scale-invariant Signal-to-Distortion Ratio (SI-SDR):}
SI-SDR~\citep{kolbaek2020loss} evaluates reconstruction fidelity while explicitly removing sensitivity to global gain mismatch by optimally projecting the estimated signal onto the subspace spanned by the reference signal, which is defined as
\begin{equation}
\text{SI-SDR}(\mathbf{s}, \hat{\mathbf{s}}) = 10 \cdot \log_{10} \left(
\frac{\| \alpha \mathbf{s} \|^2}{\| \hat{\mathbf{s}} - \alpha \mathbf{s} \|^2}
\right),
\label{eq:sisdr}
\end{equation}
where
\begin{equation}
\alpha = \frac{\langle \hat{\mathbf{s}}, \mathbf{s} \rangle}{\| \mathbf{s} \|^2}.
\label{eq:alpha}
\end{equation}
This metric evaluates reconstruction quality by measuring the energy ratio between the optimally scaled target signal and the residual distortion. Owing to its scale-invariant formulation and robustness to gain mismatch, SI-SDR has become a standard evaluation criterion for speech separation, \ac{se}, and \ac{bwe}/\ac{asr} in contemporary audio research. However, it shares the same fundamental limitation as SNR for \ac{bwe}/\ac{asr}: sample-level alignment between generated \ac{hf} content and the reference is not required for perceptual fidelity, so SI-SDR can underrate perceptually high-quality reconstructions.

\textbf{Log-Spectral Distance (LSD):}
LSD~\citep{gray1976distance} measures reconstruction quality by computing the average discrepancy between the logarithmic magnitude spectra of the reference and reconstructed signals across frequencies.
Let $\mathbf{S} \in \mathbb{C}^{T \times F}$ and $\hat{\mathbf{S}} \in \mathbb{C}^{T \times F}$ denote the \ac{stft} of $\mathbf{s}$ and $\hat{\mathbf{s}}$, respectively, where $T$ and $F$ denote the numbers of time frames and frequency bins in the spectrogram.
Their magnitude spectra are defined as $\mathbf{X} = |\mathbf{S}| \in \mathbb{R}^{T \times F}$ and $\hat{\mathbf{X}} = |\hat{\mathbf{S}}| \in \mathbb{R}^{T \times F}$. LSD is then computed as
\begin{equation}
\text{LSD}(\mathbf{s}, \hat{\mathbf{s}}) = \frac{1}{T} \sum_{t=1}^{T} 
    \sqrt{\frac{1}{F} \sum_{f=1}^{F} D_{t,f}^2},
\label{eq:lsd1}
\end{equation}
with the per-frame, per-frequency log-spectral difference
\begin{equation}
D_{t,f} = \log_{10} \left( \frac{\hat{\mathbf{X}}_{t,f}}{\mathbf{X}_{t,f}} \right).
\label{eq:lsd2}
\end{equation}
To evaluate different spectral regions separately, the \ac{lf} LSD (LSD-LF) is used to assess reconstruction quality of the preserved \ac{lf} region, 
while the \ac{hf} LSD (LSD-HF) measures the quality of the extended \ac{hf} region. They are defined as
\begin{equation}
\text{LSD-HF}(\mathbf{s}, \hat{\mathbf{s}}) = \frac{1}{T} \sum_{t=1}^{T} 
    \sqrt{\frac{1}{F_\text{HF}} \sum_{f \in \mathcal{F}_\text{HF}} D_{t,f}^2 },
\label{eq:lsdhf}
\end{equation}
\begin{equation}
\text{LSD-LF}(\mathbf{s}, \hat{\mathbf{s}}) = \frac{1}{T} \sum_{t=1}^{T} 
    \sqrt{\frac{1}{F_\text{LF}} \sum_{f \in \mathcal{F}_\text{LF}} D_{t,f}^2 },
\label{eq:lsdlf}
\end{equation}
where $\mathcal{F}_\text{HF}$ and $\mathcal{F}_\text{LF}$ denote the sets of \ac{hf} and \ac{lf} bins, with $F_\text{HF}$ and $F_\text{LF}$ representing the numbers of bins in \ac{hf} and \ac{lf} regions, respectively. Lower scores in LSD, LSD-HF, and LSD-LF imply a more similar spectrum to the target. While LSD correlates better with perceptual quality than time-domain measures like SNR, it often fails to account for the quality of reconstructed signals that are perceptually authentic but not mathematically identical to the reference spectra.

\textbf{Virtual Speech Quality Objective Listener (ViSQOL):}
ViSQOL~\citep{hines2015visqol} is a signal-based, full-reference metric that estimates perceptual quality using a spectro-temporal measure of similarity between reference and degraded signals. 
It outputs a Mean Opinion Score-Listening Quality Objective (MOS-LQO), where higher values indicate better quality. ViSQOL supports both speech and general audio: in speech mode ($16$ kHz), MOS-LQO ranges from $1$ to $5$, while in audio mode ($48$ kHz), it ranges from $1$ to $4.75$. Unlike SNR-based metrics, ViSQOL is better suited for \ac{bwe}/\ac{asr} evaluation as it reflects perceptual similarity rather than sample-level fidelity.

\textbf{Perceptual Evaluation of Speech Quality (PESQ):}
PESQ~\citep{beerends2002perceptual} is an ITU-T standardized metric that estimates perceptual speech quality by comparing reference and degraded signals. The original PESQ targets \ac{nb} speech ($300$ Hz–$3.4$ kHz) with scores ranging from $-0.5$ to $4.5$, while its \ac{wb} extension (WB-PESQ)\footnote{\url{https://www.itu.int/rec/T-REC-P.862.2}} supports $16$ kHz speech with scores between $1.04$ and $4.64$. In both cases, higher values indicate better quality. However, PESQ is designed for distortions typical of telecommunications channels (e.g., coding artifacts, packet loss) rather than \ac{hf} reconstruction, and can therefore be misleading for evaluating \ac{bwe}/\ac{asr} systems.

\textbf{Short-Time Objective Intelligibility (STOI):}
STOI~\citep{taal2010short} is an objective metric that estimates speech intelligibility by correlating short-time temporal envelopes of reference and degraded signals, producing scores between $0$ and $1$, where higher values indicate better intelligibility. Its extension, ESTOI~\citep{jensen2016algorithm}, incorporates longer temporal dependencies to better match human perception under noisy and reverberant conditions. STOI/ESTOI are most relevant for \ac{nb} speech \ac{bwe}, where intelligibility is the primary goal, but are less informative for music \ac{asr} or high-fidelity speech reconstruction, where perceptual quality is more critical.
\subsubsection{Subjective Evaluations}
\textbf{Mean Opinion Score (MOS):}
MOS~\citep{wester2015we} is a subjective evaluation method standardized by the ITU-T for assessing perceived speech and audio quality. It is computed by averaging listener ratings, typically collected on a five-point Absolute Category Rating scale, where $1$ denotes bad quality and $5$ denotes excellent quality. As a direct measure of human perception, MOS serves as a reliable benchmark and is widely used to complement objective metrics in speech and audio research. In \ac{bwe}/\ac{asr}, MOS evaluations are increasingly conducted via crowdsourcing platforms such as Amazon Mechanical Turk, while automatic MOS predictors, such as MOSNet~\citep{lo2019mosnet}, DNSMOS~\citep{reddy2021dnsmos},
NISQA~\citep{mittag2021nisqa},
and WV-MOS\footnote{\url{https://github.com/AndreevP/wvmos}}, have emerged as scalable alternatives to full listening tests.

\textbf{Preference Tests:}
\label{sec:preference}
Preference-based listening tests are subjective evaluations in which listeners compare audio samples produced by different systems and indicate their preference. In A/B tests, two audio samples are directly compared in terms of perceived quality or intelligibility, while ABX tests require listeners to identify which of two candidates is closer to a reference signal. These methods enable direct pairwise comparisons and are often used alongside MOS evaluations, providing higher sensitivity to subtle differences between high-performing \ac{bwe}/\ac{asr} systems.

\textbf{Multiple Stimuli with Hidden Reference and Anchor (MUSHRA):}
\label{sec:mushra}
The MUSHRA test\footnote{\url{https://www.itu.int/rec/R-REC-BS.1534-3-201510-I/en}} is a standardized subjective evaluation methodology designed to assess intermediate-quality audio systems.
In this test, listeners are presented with multiple stimuli, including a hidden reference and one or more anchor signals, and are asked to rate the perceived audio quality on a continuous scale from $0$ to $100$.
Compared with MOS-based evaluations, MUSHRA provides finer resolution and improved sensitivity for distinguishing perceptual differences among systems. In \ac{bwe}/\ac{asr} studies, the \ac{bl}/\ac{lr} input signal is typically used as the low-quality anchor, grounding the rating scale in the degradation being addressed.

In practice, LSD and MOS are the most consistently co-reported metrics across the \ac{bwe}/\ac{asr} literature—LSD for its direct spectral relevance to the \ac{hf} reconstruction task and MOS for its alignment with human perceptual judgment—while MUSHRA and ViSQOL are increasingly adopted in higher-fidelity music and general audio settings.

%% file: sec/5_discriminative.tex
\section{Discriminative Models}
\label{sec:dis_model}

Discriminative models learn direct mappings from \ac{bl}/\ac{lr} representations to their \ac{bb}/\ac{hr} counterparts. This section categorizes existing approaches into \ac{mlp}, \ac{rnn}, \ac{cnn}, Transformer, and Mamba-based frameworks, reflecting the progression from foundational to more advanced architectures. Table~\ref{tab:discri_arch} summarizes representative discriminative methods. Note that these categories refer to backbone designs rather than mutually exclusive modeling paradigms, and similar architectures may also be used within generative systems.

\begin{table}[htbp]
\centering
\renewcommand{\arraystretch}{1.25}
\setlength{\tabcolsep}{3pt}
\caption{\textbf{Taxonomy of discriminative models for \ac{bwe}/\ac{asr}} according to their architectural design. Some methods span multiple categories as they combine architectures (e.g., hybrid CNN-RNN designs).}
\vspace{0pt}
\label{tab:discri_arch}

\small
\begin{tabular*}{\linewidth}{@{}c@{\hspace{20pt}}c@{}}
\toprule
\textbf{Architecture} & \textbf{Literature} \\
\midrule

MLP &
\makecell[c]{
\citet{li2015dnn}, \citet{li2015deep}, \citet{wang2015speech},
\citet{gu2015restoring},
\citet{liu2015novel},\\
\citet{abel2016artificial}, \citet{abel2017artificial}, 
\citet{abel2018simple},
} \\
\addlinespace[5pt]

RNN &
\makecell[c]{
\citet{liu2016novel}, \citet{gu2016speech}, \citet{schmidt2018blind}, 
HRNN~\citep{ling2018waveform},\\
TFiLM~\citep{birnbaum2019temporal},
RDPN~\citep{hou2020speaker}, 
2S-BWE~\citep{lin2021two},\\
DCCRN~\citep{edraki2024speaker},
} \\
\addlinespace[5pt]

CNN &
\makecell[c]{
AudioUNet~\citep{kuleshov2017audio}, 
DCNN~\citep{gu2017waveform},
TFNet~\citep{lim2018time},\\
\citet{schmidt2018blind},
TFiLM~\citep{birnbaum2019temporal},
RDPN~\citep{hou2020speaker},\\
\citet{dong2020time},
\citet{hao2020time},
AECNN~\citep{wang2020time,wang2021towards},\\
AFiLM~\citep{rakotonirina2021self},
2S-BWE~\citep{lin2021two},
TUNet~\citep{nguyen2022tunet}, \\
EP-WUN~\citep{lin2023noise},
\citet{li2023restoration},
DCCRN~\citep{edraki2024speaker},\\
CLASS~\citep{kim2024class},
CTFT-Net~\citep{tamiti2025high},
} \\
\addlinespace[5pt]

Transformer &
\makecell[c]{
AFiLM~\citep{rakotonirina2021self},
TUNet~\citep{nguyen2022tunet}, 
TRAMBA~\citep{sui2024tramba},\\
CLASS~\citep{kim2024class},
DAC-NRBWE~\citep{liu2025neural},
CTFT-Net~\citep{tamiti2025high}
} \\
\addlinespace[5pt]

Mamba &
\makecell[c]{
TRAMBA~\citep{sui2024tramba}
} \\

\bottomrule
\end{tabular*}
\end{table}

\subsection{Multilayer Perceptron (MLP)}
Early studies~\citep{li2015dnn,liu2015novel,wang2015speech,gu2015restoring,li2015deep} formulate \ac{bwe} as a high-dimensional nonlinear mapping problem and employ \ac{mlp}s to model the complex relationship between \ac{lb} spectral envelopes and \ac{ub} components. A common strategy is to first perform unsupervised pre-training using \ac{rbms}~\citep{ackley1985learning} to initialize the \ac{dnn} weights and then discriminative fine-tuning with the \ac{mmse} criterion to estimate \ac{ub} spectral parameters. The \ac{ub} phase is typically recovered via inverse mirroring of the phase from the \ac{bl} signal. Abel \textit{et al.}~\citep{abel2016artificial,abel2017artificial,abel2018simple} compare \ac{hmm}-\ac{gmm}, \ac{dnn}-\ac{hmm} hybrid, and \ac{mlp}-only approaches for \ac{bwe}, finding that direct \ac{mlp}-based cepstral prediction with \ac{lb} phase-copying achieves more accurate \ac{ub} energy estimation and improved perceptual quality at low computational cost. The \ac{mlp}-based approaches have outperformed conventional \ac{gmm}-based methods across both objective and subjective evaluations.

\subsection{Recurrent Neural Network (RNN)}
\ac{rnn}s constitute an important model family for audio signal processing because they are designed to capture both short-range and long-range temporal dependencies. \citet{gu2016speech} enhance \ac{bwe} by incorporating linguistic information through bottleneck features extracted from an \ac{mlp}-based \ac{hmm} state classifier, providing a compact representation of both acoustic and phonetic cues. To improve the reduction of discontinuities from frame-independent mapping, this method employs a deep \ac{rnn} with stacked \ac{lstm} layers, outperforming both \ac{gmm}-based and \ac{mlp}-based approaches. \citet{liu2016novel} propose a two-stage \ac{bwe} approach that uses a \ac{blstm} to estimate the \ac{ub} \ac{lps} from \ac{lb} acoustic features, followed by exemplar-based sparse refinement to mitigate over-smoothing before combining with \ac{lb} components. HRNN~\citep{ling2018waveform} employs recurrent layers at multiple temporal resolutions to efficiently capture long-range dependencies. It directly predicts the \ac{bb} waveform from the \ac{bl} input and supports conditioning using bottleneck features from a \ac{dnn}-based state classifier.

Subsequent works~\citep{schmidt2018blind,birnbaum2019temporal,hou2020speaker,lin2021two,edraki2024speaker}, further explore hybrid \ac{cnn}-\ac{rnn} architectures that combine convolutional feature extraction with recurrent temporal modeling to improve \ac{bwe}/\ac{asr} performance. \citet{schmidt2018blind} present a \ac{bwe} system that estimates \ac{ub} spectral envelope from \ac{lb} power spectra using a regressive \ac{cnn}-\ac{lstm} architecture, which improves speech quality without additional algorithmic delay and can be integrated into modern speech and audio codecs. \citet{birnbaum2019temporal} introduce Temporal Feature-Wise Linear Modulation (TFiLM), a mechanism that captures long-range dependencies by modulating convolutional activations with contextual information extracted by an \ac{rnn}. This hybrid design effectively expands the receptive field of convolutional sequence models with minimal computational overhead, achieving superior \ac{asr} performance over the \ac{mlp}-based model. \citet{hou2020speaker} combine acoustic features with a low-dimensional feature vector that summarizes speaker-specific characteristics and phonetic posteriorgrams (PPGs) and propose a residual dual-path network (RDPN) for \ac{bwe}. The RDPN segments the input sequence into overlapping chunks and applies \ac{blstm} layers along intra-chunk and inter-chunk dimensions to capture utterance-level temporal dependencies while mitigating gradient-vanishing issues. 
\citet{lin2021two} introduce a two-stage \ac{bwe} (2S-BWE) framework that leverages the complementary strengths of frequency-domain and time-domain modeling to reconstruct high-fidelity speech. The approach employs a spectrogram-based network—utilizing either a convolutional recurrent network (CRN)~\citep{tan2018convolutional} or a temporal convolutional network (TCN)~\citep{bai2018empirical}—to estimate the \ac{hf} spectral envelope, followed by a Wave-U-Net-based~\citep{stoller2018wave} refinement stage trained with a multi-resolution STFT loss~\citep{defossez2020real} to improve phase consistency and enhance temporal details.
\citet{edraki2024speaker} investigate \ac{bwe} for bone-conducted speech using a deep complex convolutional-recurrent network (DCCRN)~\citep{hu2020dccrn}, which comprises a pair of complex \ac{2d} convolutional encoder-decoder and stacked \ac{lstm} layers. To improve generalization in this highly speaker-dependent setting, the authors introduce a bottleneck module that disentangles speaker-specific characteristics from speech content, therefore enabling the architecture to adapt more effectively to unseen speakers. 

\subsection{Convolutional Neural Network (CNN)}
\ac{cnn}-based \ac{bwe}/\ac{asr} models can be broadly grouped into three categories: direct feed-forward \acp{cnn}~\citep{schmidt2018blind}, including dilated variants~\citep{gu2017waveform}; autoencoder-based \acp{cnn}~\citep{dong2020time,wang2020time,hou2020speaker,wang2021towards,edraki2024speaker}; and U-Net-based \acp{cnn}~\citep{kuleshov2017audio,lim2018time,birnbaum2019temporal,lin2021two,rakotonirina2021self,nguyen2022tunet,lin2023noise,li2023restoration,kim2024class,tamiti2025high}. Earlier studies employed direct feed-forward architectures that process features through sequential convolutions or autoencoder structures that compress them into bottleneck representations before reconstruction.
DCNN~\citep{gu2017waveform} employs WaveNet-inspired causal and non-causal dilated convolutions for waveform-level \ac{bwe}. The causal variant supports real-time processing, whereas the non-causal variant uses future context for offline reconstruction. Unlike WaveNet, both directly estimate the \ac{bb} waveform without autoregressive output feedback.
\citet{dong2020time} propose a \ac{bwe} method based on a time-frequency network whose dual branches are constructed from an autoencoder architecture enhanced with channel-attention and non-local modules to achieve richer channel-wise and spatial feature modeling. By jointly predicting waveform-level phase in the time branch and spectral-magnitude information in the frequency branch, the model enables concurrent phase-magnitude modeling and avoids invalid \ac{stft} reconstruction. 
\citet{wang2020time,wang2021towards} employ an autoencoder architecture called AECNN~\citep{pandey2019new} and introduce cross-domain training objectives for \ac{ssr}, combining time-domain waveform losses with frequency-domain \ac{stft} losses to achieve balanced reconstruction of temporal and spectral structures. In addition, the transposed-convolution upsampling layers in AECNN are replaced with subpixel layers to mitigate artifacts and improve computational efficiency.
As discussed in the previous section, \citet{hou2020speaker,edraki2024speaker} integrate \ac{lstm} modules into an autoencoder-based \ac{bwe} framework. Although the above methods are effective, plain feed-forward \ac{cnn}s are limited by single-scale receptive fields, while autoencoder-based models may suffer from information loss due to heavy bottleneck compression. 

\begin{figure}[t]
\centering
\includegraphics[width=0.75\linewidth]{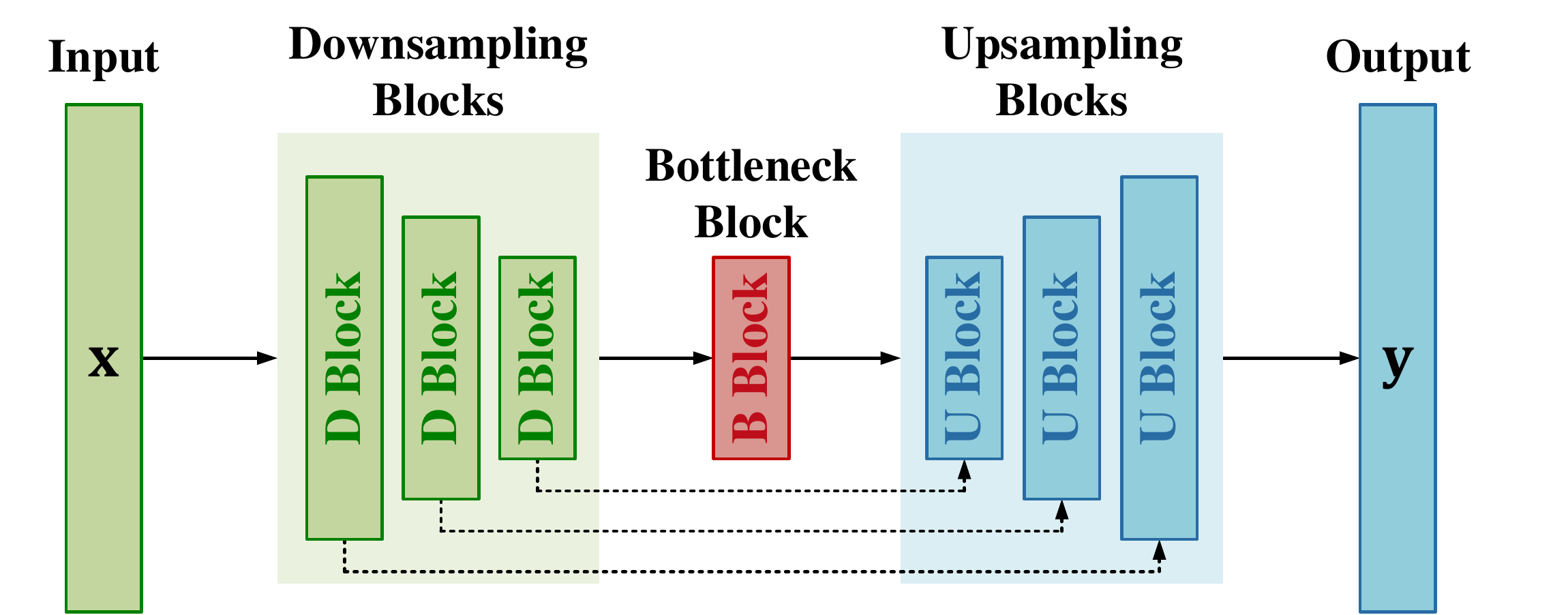}
\caption{\textbf{The U-Net architecture.} It employs a symmetric encoder-decoder structure with multi-scale skip connections that align the corresponding stages, while the bottleneck block forms the most compact latent representation.}
\label{fig:unet}
\end{figure}

To address these limitations, later designs have evolved toward U-Net architectures that employ multi-scale encoder-decoder pathways with aligned skip connections, as illustrated in Fig.~\ref{fig:unet}. Through progressive downsampling and upsampling across multiple resolutions, the multi-scale encoder-decoder pathways capture local details while integrating long-range contextual information, whereas the aligned skip connections facilitate direct cross-resolution feature propagation, mitigating information loss caused by compression and facilitating more accurate reconstruction.
AudioUNet~\citep{kuleshov2017audio}, the first U-Net architecture applied to \ac{asr}, operates directly in the raw waveform and is trained end-to-end with a time-domain \ac{mse} loss. In addition, the model incorporates 1-D subpixel-shuffling layers, enabling high-fidelity reconstruction and reducing artifacts relative to conventional upsampling across multiple upscaling ratios. \citet{lim2018time} introduce an end-to-end Time-Frequency Network (TFNet), in which both time-domain and frequency-domain branches follow the architecture design patterns from AudioUNet~\citep{kuleshov2017audio}. By jointly optimizing the time-domain waveform and frequency-domain magnitude representations through a spectral fusion layer, the model implicitly models phase in the time domain and the magnitude in the frequency domain. This cross-domain formulation yields significantly higher-quality reconstructions than prior single-domain \ac{asr} approaches. \citet{li2023restoration} propose a U-Net-based bone-conducted speech \ac{bwe} framework in which a shifted window-based attention mechanism~\citep{liu2021swin} is integrated into each residual 1-D convolutional encoder-decoder block, enabling effective modeling of long-range temporal dependencies. In addition, the method replaces the conventional time-domain \ac{mse} loss with a multi-scale mel-spectral energy distance objective, resulting in improved \ac{hf} restoration performance. 
EP-WUN~\citep{lin2023noise} integrates a Wave-U-Net backbone~\citep{stoller2018wave} with a triplet loss-based embedding polishing stage to simultaneously perform noise-robust \ac{bwe} and latent-space signal purification.
More recent works~\citep{rakotonirina2021self,nguyen2022tunet,kim2024class,sui2024tramba,tamiti2025high,liu2025neural} integrate Transformers~\citep{vaswani2017attention} into U-Net architectures, which are analyzed in the following section.

\subsection{Transformer}
Transformers~\citep{vaswani2017attention,wolf2020transformers,kwon2024poster,yao2025generality,yang2026fsd50k,song2026channel} have increasingly been adopted for \ac{bwe}/\ac{asr} due to their ability to capture long-range temporal and spectral dependencies through a global self-attention mechanism. Unlike convolutional models, which are constrained by local receptive fields, Transformers can model interactions between distant time-frequency regions, which are crucial for restoring \ac{hf} harmonics and capturing phoneme-level context. Transformer-based discriminative models~\citep{rakotonirina2021self,nguyen2022tunet,kim2024class,tamiti2025high,liu2025neural} are commonly embedded within U-Net structures, where self-attention modules replace or augment convolutional blocks at the bottleneck or intermediate feature levels, enabling global context modeling while retaining the multi-scale benefits of U-Net feature fusion. AFiLM~\citep{rakotonirina2021self} is among the first to introduce Transformers into \ac{asr}, replacing \ac{rnn}-based conditioning with a self-attention mechanism through an Attention-based Feature-Wise Linear Modulation layer. AFiLM enables the model to capture long-range temporal dependencies beyond the limits of local convolutions, which leads to faster training and improved reconstruction accuracy. Subsequently, \citet{nguyen2022tunet} propose Transformer-aided U-Net (TUNet), which employs a lightweight 1-D U-Net backbone with a Transformer encoder at the bottleneck, enabling global-dependency modeling while maintaining computational efficiency through aggressive time-downsampling. While the Transformer refines the compressed latent representation, the U-Net handles hierarchical feature extraction and reconstruction, resulting in a compact yet highly effective \ac{bwe} architecture. \citet{kim2024class} introduce a continual learning approach for \ac{ssr} (CLASS), which extends TUNet~\citep{nguyen2022tunet} by integrating self-supervised pretraining and continual learning to prevent catastrophic forgetting when transitioning from masked speech modeling to \ac{ssr}. This strategy preserves prior representational knowledge while improving reconstruction performance and cross-condition generalization.
\citet{tamiti2025high} propose a complex-domain time-frequency network (CTFT-Net) that incorporates a conformer bottleneck into the U-Net to jointly recover the magnitude and phase from complex spectrograms. The complex global attention block (CGAB) effectively captures long-range time-frequency dependencies, resulting in superior \ac{hf} reconstruction for \ac{ssr}, particularly under extreme upsampling conditions. \citet{liu2025neural} propose a neural codec-based framework for noise-robust \ac{bwe} in which a conformer-based neural network learns a discriminative mapping from noisy codec embeddings to clean codebook indices. The predicted discrete tokens are then decoded by a pre-trained \ac{nac} to reconstruct the enhanced \ac{bb} speech, achieving competitive performance among noise-robust \ac{bwe} methods.

\subsection{Mamba}
Mamba~\citep{gu2024mamba} is a sequence modeling architecture based on selective state-space models (SSMs) that enables efficient modeling of long-range dependencies with linear computational complexity in sequence length. Compared with \ac{cnn}s, which rely on finite receptive fields or deep stacking to capture long-term context, Mamba naturally models global temporal dependencies through its state-space formulation. In contrast to Transformer-based models, whose self-attention incurs quadratic complexity with respect to sequence length, Mamba achieves scalable and memory-efficient sequence modeling while maintaining strong expressive power. TRAMBA~\citep{sui2024tramba} presents a hybrid Transformer-Mamba \ac{ssr} framework specifically for deployment on mobile and wearable platforms. By pretraining on large-scale speech datasets and fine-tuning with only a small amount of user-specific bone-conduction data, TRAMBA effectively addresses data scarcity and reduces the burden of modality-specific data collection. Extensive evaluations show that TRAMBA achieves substantial improvements in speech quality and intelligibility while maintaining a compact model size and enabling highly efficient inference. Furthermore, real-system integration and user studies demonstrate that TRAMBA delivers robust \ac{ssr} performance in noisy environments and significantly improves wearable battery life by reducing sensing and data transmission requirements.
Beyond purely discriminative settings, Mamba-based designs have also been recently adopted within generative frameworks: Wave-U-Mamba~\citep{lee2025wave}, VM-ASR~\citep{zhang2025vm}, EDNet~\citep{kwak2026ednet}, HCGAN~\citep{gao2026harmonic}, and SEMamba++~\citep{lee2026semambaplusplus} integrate Mamba blocks into GAN generators (in Section~\ref{sec:gen model}), and VQ-Diffusion~\citep{fang2025vector} employs a Mamba-2 backbone within a discrete diffusion model for token-level \ac{bwe}. This cross-paradigm adoption underscores the versatility of state-space sequence modeling as an alternative to self-attention across \ac{bwe}/\ac{asr} architectures.

Across the discriminative paradigm, the progression from \ac{mlp}s to \ac{rnn}s, \ac{cnn}s, Transformers, and Mambas reflects a consistent drive toward richer temporal context, greater spectral detail, and more efficient computation. While these models achieve strong objective performance, they share a fundamental limitation: trained with distance-based losses, they tend to produce over-smoothed reconstructions that lack the fine stochastic \ac{hf} structure of natural audio. This motivates the shift to generative modeling frameworks, surveyed in the following section.

%% file: sec/6_generative_final.tex
\section{Generative Models}
\label{sec:gen model}

Generative models for \ac{bwe}/\ac{asr} move beyond direct point estimation by learning probabilistic or perceptually motivated reconstruction objectives. While \ac{ar}, \ac{vae}, diffusion, flow, and bridge models can support stochastic conditional generation, many GAN-based systems remain deterministic at inference despite using adversarial training. This section categorizes the literature by modeling paradigm, including \ac{ar}, \ac{vae}, \ac{gan}, diffusion, flow, and bridge models, with representative approaches summarized in Table~\ref{tab:gen_arch}.

\begin{table}[h]
\centering
\renewcommand{\arraystretch}{1.25}
\setlength{\tabcolsep}{3pt}
\caption{\textbf{Taxonomy of generative approaches for \ac{bwe}/\ac{asr}} according to their modeling paradigms.}
\vspace{0pt}
\label{tab:gen_arch}

\small
\begin{tabular*}{\linewidth}{@{}c@{\hspace{14pt}}c@{}}
\toprule
\textbf{Method} & \textbf{Literature} \\
\midrule

AR &
\makecell[c]{
Parallel-WaveNet~\citep{wang2018speech},
WaveNet~\citep{gupta2019speech},
HP-codecX~\citep{ginies2025harmonic}
} \\
\addlinespace[5pt]

VAE &
\makecell[c]{
CVAE-BWE~\citep{bachhav2020artificial},
TFDFE-ASR~\citep{tian2024time}
} \\
\addlinespace[5pt]

GAN &
\makecell[c]{
\citet{li2018speech},
\citet{eskimez2019speech},
\citet{li2019speech},
CycleGAN~\citep{haws2019cyclegan},\\
\citet{sautter2019artificial},
MU-GAN~\citep{kim2019bandwidth},
MfNet~\citep{hao2020time},
\citet{su2021bandwidth},\\
Stream-SEANet~\citep{li2021real},
BEHM-GAN~\citep{moliner2022behm},
NVSR~\citep{liu2022neural},\\
VoiceFixer~\citep{liu2022voicefixer},
EBEN~\citep{hauret2023eben},
Dual-CycleGAN~\citep{yoneyama2023nonparallel},\\
HiFi++~\citep{andreev2023hifiplusplus},
BBWEXNet~\citep{gomez2023low},
mdctGAN~\citep{shuai2023mdctgan},\\
MLPStreams~\citep{soltanmohammadi2023low},
AERO~\citep{mandel2023aero},
\citet{li2023two},\\
MS-BWE~\citep{lu2024multi},
Fre-Painter~\citep{kim2024audio},
AP-BWE~\citep{lu2024towards},\\
SDNet~\citep{yang2024sdnet},
BAE-Net~\citep{yu2024bae},
ICCRN-EBEN~\citep{bao2025frequency}\\
~\citet{kataria2024time},
SwinSRGAN~\citep{yuan2025swinsrgan},
NLDSI-BWE~\citep{tamiti2025nldsi},\\
SFNet~\citep{dai2025sfnet},
Wave-U-Mamba~\citep{lee2025wave},
VM-ASR~\citep{zhang2025vm},\\
HiFi-SR~\citep{zhao2025hifi},
HWB-Net~\citep{liu2025hwb},
HCGAN~\citep{gao2026harmonic},\\
SEMamba++~\citep{lee2026semambaplusplus},
EDNet~\citep{kwak2026ednet},
\citet{gao2026real},\\
} \\
\addlinespace[5pt]

Diffusion &
\makecell[c]{
NU-Wave~\citep{lee2021nu},
UNIVERSE~\citep{serra2022universal},
NU-Wave2~\citep{han2022nu},\\
\citet{lemercier2023analysing},
UDM+~\citep{yu2023conditioning},
UNIVERSE++~\citep{scheibler2024universal},\\
SWiBE~\citep{lin2024swibe},
AudioSR~\citep{liu2024audiosr},
BABE~\citep{moliner2024blind},\\
FlashSR~\citep{im2025flashsr},
VQ-Diffusion~\citep{fang2025vector},
IS-Diffusion~\citep{jin2026inference},\\
FastWave~\citep{kuznetsov2026fastwave}
} \\
\addlinespace[5pt]

Flow &
\makecell[c]{
WSRGlow~\citep{zhang2021wsrglow},
GSFM~\citep{ku2025generative},
Stream FM~\citep{welker2025real},\\
FLowHigh~\citep{yun2025flowhigh},
UniverSR~\citep{choi2025universr},
SAGA-SR~\citep{im2026saga},\\
CodecFlow~\citep{zhang2026codecflow},
Real-time Flow~\citep{hsieh2026towards},\\
DSC-FLowHigh~\citep{hernandez2026single},
LatentFlowSR~\citep{liu2026latentflowsr}
},
\\
\addlinespace[5pt]

Bridge &
\makecell[c]{
A2SB~\citep{shihaudio},
Bridge-SR~\citep{li2025bridge},
AudioLBM~\citep{li2025audio}
} \\

\bottomrule
\end{tabular*}
\end{table}

\subsection{Autoregressive (AR) Model}
WaveNet~\citep{van2016wavenet} is a fully probabilistic \ac{ar} model originally developed for audio generation tasks such as multi-speaker speech synthesis, text-to-speech (TTS), and musical audio modeling, where each sample is predicted from all previous samples. This framework is then extended to \ac{bwe}/\ac{asr}~\citep{wang2018speech,gupta2019speech} by adopting a conditional WaveNet to model the conditional distribution of a \ac{bb}/\ac{hr} waveform $\mathbf{y} = [y[1], \ldots, y[T]]^\top$ given a \ac{bl}/\ac{lr} waveform $\mathbf{x}$. The formulation is given as
\begin{equation}
\label{eq:conditional_wavenet}
p(\mathbf{y} \mid h(\mathbf{x})) = \prod_{t=1}^{T} p\left(y[t] \mid y[1], \ldots, y[t-1], h(\mathbf{x})\right),
\end{equation}
where the conditioning term $h(\mathbf{x})$ can be either a raw waveform $h(\mathbf{x}) = \mathbf{x}$ or a log-mel spectrogram
\begin{equation}
\label{eq:transform}
h(\mathbf{x}) = \log \left({\text{MelFilter}}\left(|\text{STFT}(\mathbf{x})|\right)\right).
\end{equation}
The log-mel spectrogram conditioning has been shown to outperform raw waveform conditioning~\citep{wang2018speech,gupta2019speech} and is widely recognized as an effective intermediate representation for high-fidelity audio generation~\citep{shen2018natural}.

\begin{figure}[t]
\centering
\includegraphics[width=0.75\linewidth]{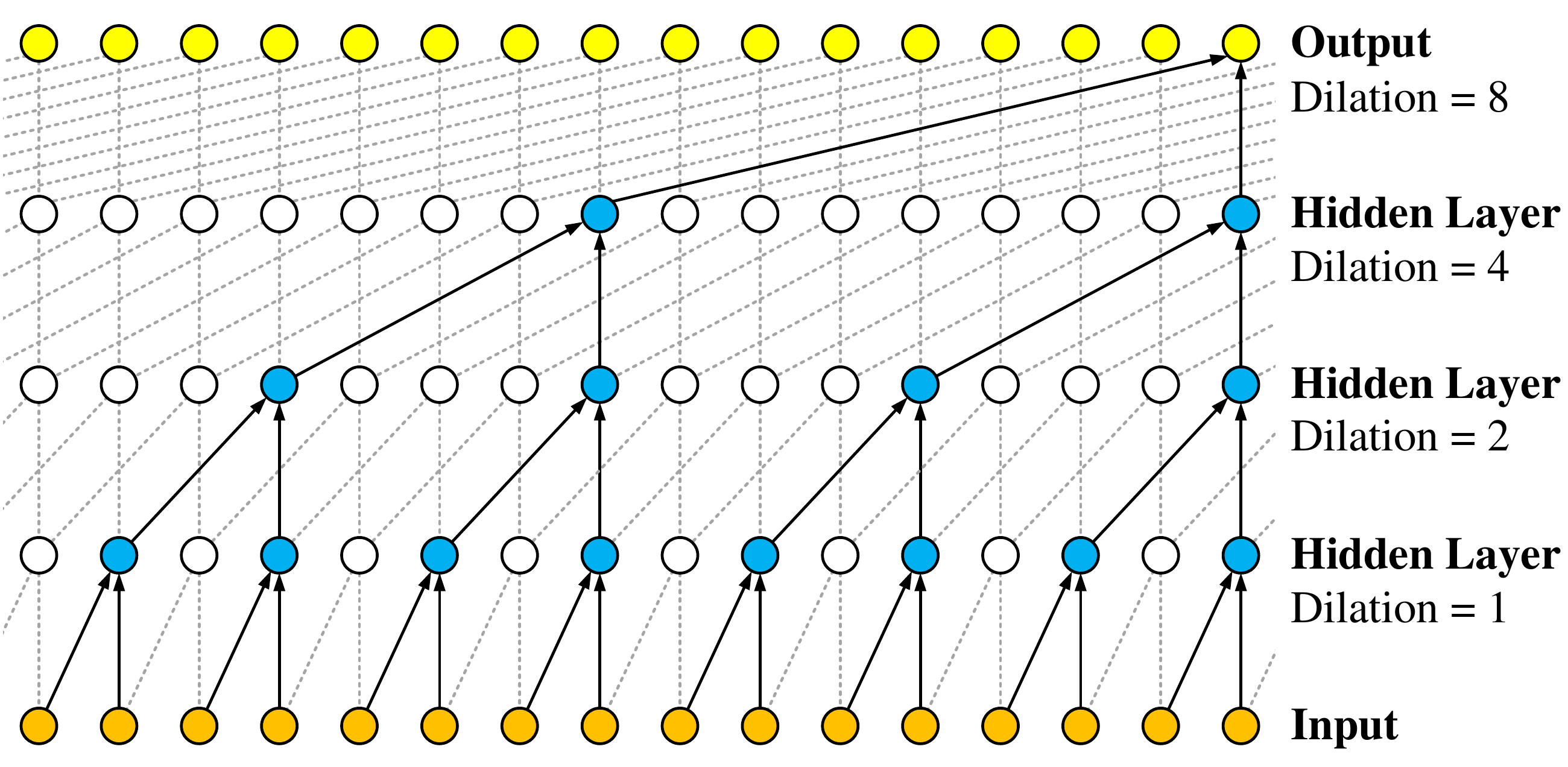}
\caption{\textbf{Visualization of a stack of dilated causal convolutional layers.} Dilated causal convolutions with dilation factors of $1$, $2$, $4$, and $8$ are shown, where dilation specifies the spacing between consecutive filter taps, allowing the temporal receptive field to grow exponentially while preserving causality.}
\label{fig:dilated_cnn}
\end{figure}

WaveNet is built on causal convolutions that enforce temporal causality by conditioning each sample \( y[t] \) only on past samples, while enabling parallel computation during training and efficient modeling of long sequences.
To expand temporal context without excessively deep stacks or large kernels, WaveNet employs dilated causal convolutions, where the dilation factor increases exponentially across layers, resulting in exponential receptive-field expansion with minimal parameter growth. Fig.~\ref{fig:dilated_cnn} depicts dilated causal convolutions for dilation = $1$,  $2$, $4$, and $8$. 
Each layer uses gated activation units~\citep{van2016conditional}, which outperform standard ReLU activations in audio modeling. The output distribution is modeled via a categorical softmax over \(\mu\)-law-companded samples with 256-level quantization~\citep{recommendation1988pulse}, and it is trained using the \ac{ce} loss. Residual and skip connections further stabilize optimization in deep architectures. 

In~\citep{gupta2019speech}, the conditioning log-mel spectrogram is processed to match the temporal resolution of the waveform and injected into WaveNet layers to guide generation. While WaveNet produces high-quality audio, its sample-by-sample generation leads to slow inference. To address this limitation, \citet{wang2018speech} propose a parallel WaveNet trained via teacher-student distillation, where a non-\ac{ar} student model approximates the \ac{ar} teacher. By leveraging inverse \ac{ar} flows, the student enables parallel generation and significantly improves inference speed. The training objective combines a \ac{kl} divergence term with a spectral-domain power loss to preserve perceptual quality.

More recently, \citet{ginies2025harmonic} propose HP-codecX, which formulates \ac{bwe} as an \ac{ar} token prediction task using a Transformer-based audio language model. The method employs a harmonic-percussive disentangled \ac{nac} to obtain structured \ac{lf} and \ac{hf} representations, enabling the Transformer to predict missing \ac{hf} tokens conditioned on the observed \ac{lf} tokens. Unlike sample-level autoregression in WaveNet, HP-codecX performs autoregressive prediction across residual vector quantization (RVQ) codebook stages, followed by neural codec decoding to reconstruct the full-band audio.


\subsection{Variational Autoencoder (VAE)}

\begin{figure}[t]
\centering
\includegraphics[width=0.75\linewidth]{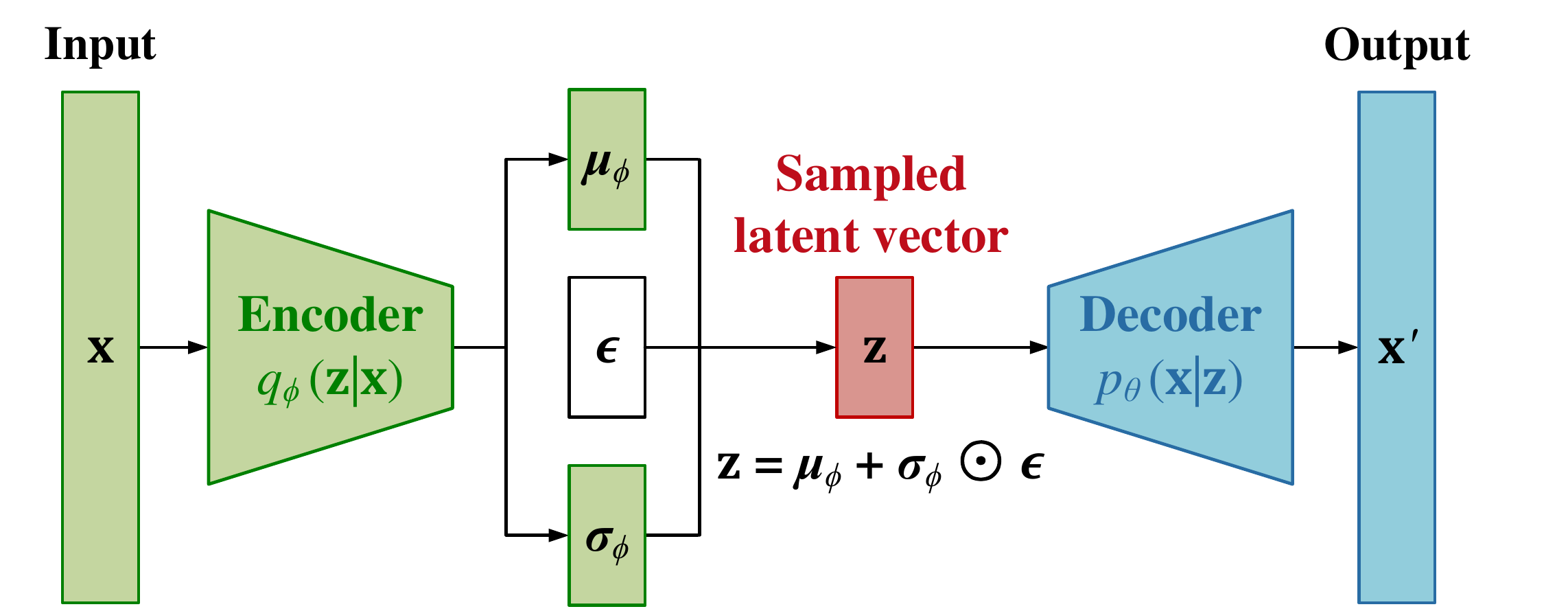}
\caption{\textbf{Architecture of an unconditional VAE,} where an encoder infers the latent distribution $\mathbf{z}$ parameterized by mean $\boldsymbol{\mu}_{\phi}$ and variance $\boldsymbol{\sigma}_{\phi}$, and a decoder reconstructs the input signal $\mathbf{x}$ via latent sampling using the reparameterization trick.}
\label{fig:vae}
\end{figure}

An unconditional variational autoencoder (VAE)~\citep{doersch2016tutorial,chen2025generating} models the generation of an input signal $\mathbf{x} \in \mathbb{R}^{T}$ through a latent variable $\mathbf{z} \in \mathbb{R}^D$. The generative process is formulated as
\begin{equation}
\label{eq:vae_prob}
p_{\theta}(\mathbf{x}) = \int p_{\theta}(\mathbf{x}\mid \mathbf{z}) \, p_{\theta}(\mathbf{z}) \, d\mathbf{z},
\end{equation}
where $p_{\theta}(\mathbf{x})$ denotes the marginal likelihood of the input signal, $p_{\theta}(\mathbf{x}\mid \mathbf{z})$ represents the conditional likelihood that maps the latent representation to the input, and $p_{\theta}(\mathbf{z})$ is the prior distribution over the latent variable, which is typically chosen as a standard Gaussian distribution to regularize the latent space. 
During training, an encoder network $q_{\phi}(\mathbf{z}\mid \mathbf{x})$ approximates the intractable posterior over latent variables, while a decoder $p_{\theta}(\mathbf{x}\mid \mathbf{z})$ reconstructs the input from the latent representation. The model parameters $\theta$ and $\phi$ are learned by maximizing the evidence lower bound (ELBO) function $\mathcal{L}(\theta,\phi;\mathbf{x})$:
\begin{equation}
\label{eq:vae_elbo}
\begin{aligned}
\log p_{\theta}(\mathbf{x}) &\ge\; \mathcal{L}(\theta,\phi;\mathbf{x}) = \mathbb{E}_{q_{\phi}(\mathbf{z}\mid \mathbf{x})}\!\left[\log p_{\theta}(\mathbf{x}\mid\mathbf{z})\right] - D_{\mathrm{KL}}\!\left(q_{\phi}(\mathbf{z}\mid \mathbf{x}) \,\Vert\, p_{\theta}(\mathbf{z})\right),
\end{aligned}
\end{equation}
where $D_{\mathrm{KL}}(\cdot\Vert\cdot)$ denotes the KL divergence. 
The first term of $\mathcal{L}(\theta,\phi;\mathbf{x})$ encourages accurate reconstruction of the input signal, while the second term regularizes the latent distribution to match the prior. 
Sampling of the latent variable is enabled by the reparameterization trick
\begin{equation}
\label{eq:vae_reparam_z}
\mathbf{z} = \boldsymbol{\mu}_{\phi} + \boldsymbol{\sigma}_{\phi} \odot \boldsymbol{\epsilon},
\end{equation}
where $\boldsymbol{\mu}_{\phi} \in \mathbb{R}^D$ and $\boldsymbol{\sigma}_{\phi} \in \mathbb{R}^D$ denote the mean and standard deviation predicted by the encoder network, respectively, $\boldsymbol{\epsilon} = [\epsilon^{(1)}, \ldots, \epsilon^{(D)}]^\top \in \mathbb{R}^D$ is an auxiliary noise vector with independent standard normal entries
\begin{equation}
\label{eq:vae_reparam_z2}
\epsilon^{(d)} \overset{\mathrm{i.i.d.}} \sim \mathcal{N}(0,1), \quad d = 1, \ldots, D,
\end{equation}
and $\odot$ denotes element-wise multiplication.
Unlike \ac{ar} models which generate samples sequentially, \acp{vae} reconstruct the entire signal $\mathbf{x}$ in a single forward pass, enabling efficient and fully parallel inference. Fig.~\ref{fig:vae} depicts the architecture of an unconditional \ac{vae}. In contrast, a \ac{cvae}~\citep{sadeghi2020audio,kim2021conditional} extends this framework by incorporating additional conditioning information into the generative process, enabling controllable generation based on the provided conditions.

\citet{tian2024time} propose a time-frequency fusion-based \ac{asr} framework that jointly leverages waveform and spectrogram representations. Their approach combines a \ac{vae}-based time-domain module, a U-Net-based spectrogram module, and an attention-based fusion mechanism to integrate complementary information from both domains. This design preserves both temporal and spectral characteristics and demonstrates strong robustness across acoustic domains, including underwater scenarios.
\citet{bachhav2020artificial} propose a \ac{cvae}-based \ac{bwe} method that models the conditional distribution of \ac{bb} speech given the \ac{bl} input features. By incorporating adversarial learning into the \ac{cvae} framework, the approach improves both perceptual speech quality and downstream speech recognition performance.

\subsection{Generative Adversarial Network (GAN)}
\acp{gan} constitute one of the most prominent families of generative models, demonstrating substantial success across image domains, including text-to-image synthesis~\citep{reed2016generative}, image SR~\citep{bulat2018learn}, and image style transfer~\citep{azadi2018multi}.  Motivated by these advances, \acp{gan} have been increasingly adopted in audio processing applications, including \ac{se}~\citep{cao2022cmgan}, audio synthesis~\citep{engel2019gansynth}, voice conversion~\citep{chen2022controlvc}, audio source separation~\citep{tan2026dtt}, and \ac{bwe}/\ac{asr}~\citep{eskimez2019adversarial,su2021bandwidth,lee2025wave}. Within the \ac{bwe}/\ac{asr} paradigm, a generator network $G$ reconstructs the \ac{bb}/\ac{hr} feature representation or waveform from the corresponding \ac{bl}/\ac{lr} input $\mathbf{x}$, while a discriminator network $D$ is trained to distinguish between the generated output $\mathbf{\hat{y}}$ and \ac{bb}/\ac{hr} target $\mathbf{y}$, thereby establishing a min-max adversarial optimization framework. This training mechanism encourages the generator to produce \ac{bb}/\ac{hr} signals that preserve natural harmonic structures and exhibit spectral characteristics consistent with authentic audio.

\ac{gan}-based \ac{bwe}/\ac{asr} approaches can be systematically categorized along three principal design dimensions: (i) generator specification, (ii) discriminator configuration, and (iii) training objective formulation. 
The generator specification addresses whether the generator operates as a standalone model or as a submodule within a composite framework, as well as the processing domain (e.g., waveform-based or spectrogram-based) and the neural architecture (e.g., \ac{mlp}, \ac{rnn}, \ac{cnn}, Transformer, or Mamba). 
Table~\ref{tab:arch_domain} provides a taxonomy of \ac{gan}-based \ac{bwe}/\ac{asr} frameworks according to the generator's processing domain and architectures.
The discriminator configuration specifies whether a single discriminator or multiple discriminators are employed; the former evaluates audio realism independently, whereas the latter operate in parallel to jointly assess perceptual fidelity across multiple representations. Finally, the training objective formulation specifies the loss function design. Collectively, these design factors govern the perceptual quality and spectral fidelity of the reconstructed signals, as well as the system robustness.

\begin{table}[t]
\centering
\renewcommand{\arraystretch}{1.25}
\setlength{\tabcolsep}{3pt}
\caption{\textbf{Taxonomy of GAN-based \ac{bwe}/\ac{asr} approaches} according to their generators' architectural design and processing domain.}
\vspace{0pt}
\label{tab:arch_domain}

\small
\begin{tabular*}{\linewidth}{@{}c@{\hspace{2pt}}c@{\hspace{7pt}}c@{}}
\toprule
\textbf{Generator} & \textbf{Waveform-based Network} & \textbf{Spectrogram-based Network} \\
\midrule

MLP & 
- & 
\makecell[c]{
\citet{li2018speech},
\citet{sautter2019artificial},\\
MLPStreams~\citep{soltanmohammadi2023low}
} \\
\addlinespace[6pt]

RNN &
\makecell[c]{
HWB-Net~\citep{liu2025hwb}
} &
\makecell[c]{
CycleGAN~\citep{haws2019cyclegan},
BAE-Net~\citep{yu2024bae},\\
SFNet~\citep{dai2025sfnet},
ICCRN-EBEN~\citep{bao2025frequency}
} \\
\addlinespace[6pt]

CNN &
\makecell[c]{
\citet{li2019speech}, \citet{su2021bandwidth},\\
MU-GAN~\citep{kim2019bandwidth},\\
MfNet~\citep{hao2020time},\\
Stream-SEANet~\citep{li2021real},\\
EBEN~\citep{hauret2023eben},\\
BBWEXNet~\citep{gomez2023low},\\
\citet{kataria2024time},\\ 
\citet{gao2026real}
}
&
\makecell[c]{
\citet{eskimez2019speech},\\
CycleGAN~\citep{haws2019cyclegan},\\
BEHM-GAN~\citep{moliner2022behm},\\
NVSR~\citep{liu2022neural},
VoiceFixer~\citep{liu2022voicefixer},\\
\citet{li2023two},
Dual-CycleGAN~\citep{yoneyama2023nonparallel},\\
HiFi++~\citep{andreev2023hifiplusplus},
AP-BWE~\citep{lu2024towards},\\
Fre-Painter~\citep{kim2024audio},
MS-BWE~\citep{lu2024multi},\\
SDNet~\citep{yang2024sdnet},
NLDSI-BWE~\citep{tamiti2025nldsi},\\
} \\
\addlinespace[6pt]

Transformer &
- &
\makecell[c]{
AERO~\citep{mandel2023aero},
mdctGAN~\citep{shuai2023mdctgan},\\
SwinSRGAN~\citep{yuan2025swinsrgan},
HiFi-SR~\citep{zhao2025hifi}
} \\
\addlinespace[6pt]

Mamba &
\makecell[c]{
Wave-U-Mamba~\citep{lee2025wave}
} &
\makecell[c]{
VM-ASR~\citep{zhang2025vm},
EDNet~\citep{kwak2026ednet},\\
SEMamba++~\citep{lee2026semambaplusplus},
HCGAN~\citep{gao2026harmonic}
} \\

\bottomrule
\end{tabular*}
\end{table}

\subsubsection{Generator}
Based on the signal representation, \ac{gan}-based approaches can be broadly categorized into waveform-based networks which operate directly on raw waveforms and spectrogram-based networks which operate on time-frequency representations. A special instance within waveform-based methods is EBEN~\citep{hauret2023eben}, which operates on \ac{pqmf}-decomposed subband signals: the lowest subband is used as input to a U-Net generator to predict all subbands, followed by \ac{pqmf} synthesis to reconstruct the \ac{bb} waveform. For spectrogram-based approaches, a variety of representations have been explored, including \ac{lsf}~\citep{li2018speech}, power or \ac{stft} magnitude spectrograms~\citep{li2018speech,eskimez2019speech}, \ac{stft} complex spectrograms~\citep{moliner2022behm,mandel2023aero,soltanmohammadi2023low,yang2024sdnet,lu2024towards,lu2024multi,yu2024bae,zhang2025vm,dai2025sfnet,tamiti2025nldsi,gao2026harmonic}, mel-spectrograms~\citep{haws2019cyclegan,liu2022neural,liu2022voicefixer,li2023two,yoneyama2023nonparallel,andreev2023hifiplusplus,kim2024audio,zhao2025hifi,lee2026semambaplusplus,kwak2026ednet}, \ac{mdct} spectrograms~\citep{shuai2023mdctgan,yuan2025swinsrgan}, and cepstral-domain features~\citep{bao2025frequency}.

Architecturally, most \ac{gan} frameworks employ a generator as a standalone model for signal reconstruction. In contrast, several works~\citep{liu2022neural,liu2022voicefixer,li2023two,andreev2023hifiplusplus,kim2024audio,zhao2025hifi} incorporate a pretrained generator—such as TFGAN~\citep{tian2020tfgan} or HiFi-GAN~\citep{kong2020hifi}—as a submodule within a larger architecture. In the first category, an earlier study~\citep{li2018speech} employs a simple four-layer MLP generator, while subsequent approaches evolve toward recurrent architectures and convolutional designs, including autoencoder and U-Net architectures. More recent approaches adopt other high-capacity convolutional backbones, such as ConvNeXt~\citep{lu2024towards,lu2024multi,tamiti2025nldsi} and Deep Complex Networks~\citep{moliner2022behm}. In addition, Transformer-based U-Net generators leverage self-attention to capture global temporal-spectral dependencies in the spectrogram domain. Specifically, mdctGAN~\citep{shuai2023mdctgan} introduces a Transformer bottleneck stack that performs global attention on frequency-consistent features, AERO~\citep{mandel2023aero} incorporates a Frequency Transformer Block before each encoder stage, and SwinSRGAN~\citep{yuan2025swinsrgan} adopts multiple Residual Swin Transformer Blocks in the bottleneck. More recently, Mamba-based generators propose state-space sequence modeling for efficient long-context reconstruction. Wave-U-Mamba~\citep{lee2025wave} inserts Mamba blocks throughout a waveform U-Net to efficiently model long-range temporal structure, while VM-ASR~\citep{zhang2025vm} applies VMamba-style Visual State Space blocks~\citep{liu2024vmamba} within a dual-stream complex-spectrogram U-Net to better capture global-local spectro-temporal features and improve harmonic reconstruction. EDNet~\citep{kwak2026ednet} adopts a dual-stream architecture where Gating Mamba modules integrate magnitude masking and mapping to guide phase reconstruction through long-range temporal–spectral modeling.
HCGAN~\citep{gao2026harmonic} employs a dual-branch generator consisting of a U-Net-based spectral subgenerator with a Mamba bottleneck and a dedicated harmonic branch that utilizes temporal-harmonic modules to estimate \ac{hf} structures for final spectral-harmonic fusion.
SEMamba++~\citep{lee2026semambaplusplus} adopts an encoder-decoder architecture, where the bottleneck employs multi-resolution time-frequency dual-processing (TFDP) blocks with Time Mamba layers for temporal modeling and a Fourier-based frequency module to capture global, local, and periodic spectral patterns.

In the second category, a complementary design incorporates a pretrained vocoder generator as a submodule within the overall architecture. NVSR~\citep{liu2022neural} and VoiceFixer~\citep{liu2022voicefixer} adopt a two-stage \ac{ssr} pipeline, in which a ResUNet-based analysis network~\citep{kong2021decoupling} first predicts the \ac{hr} mel-spectrogram from its \ac{lr} counterpart, followed by a pretrained TFGAN vocoder~\citep{tian2020tfgan} that synthesizes the \ac{hr} waveform. As errors may propagate across stages, NVSR~\citep{liu2022neural} introduces a \ac{lf} replacement step to preserve \ac{lb} fidelity. 
\citet{li2023two} propose another two-stage framework in which the first stage employs a U-Net-based generator with adversarial training to perform \ac{bwe} on bone-conducted speech in the log-mel spectrogram domain. In the second stage, a DiffWave vocoder~\citep{kong2020diffwave} is trained separately to reconstruct high-fidelity waveforms from the extended spectral features.
HiFi++~\citep{andreev2023hifiplusplus} further employs a staged generator architecture: SpectralUNet enhances the mel-spectrogram, a HiFi-GAN backbone~\citep{kong2020hifi} generates the waveform, and WaveUNet and SpectralMaskNet subsequently refine it in the time and frequency domains, respectively. Fre-Painter~\citep{kim2024audio} integrates a pretrained masked-autoencoder encoder~\citep{huang2022masked} with a HiFi-GAN vocoder~\citep{kong2020hifi}, leveraging masking-based training to enhance robustness across varying sampling rates.
HiFi-SR~\citep{zhao2025hifi} employs a Transformer-convolutional generator that encodes \ac{lr} mel-spectrograms via a hybrid MossFormer-recurrent network~\citep{zhao2025mossformer2} to capture long-range dependencies, followed by HiFi-GAN-based waveform synthesis with joint optimization.

\subsubsection{Discriminator}
Early \acp{gan}~\citep{li2018speech,eskimez2019speech,li2019speech,sautter2019artificial,hao2020time} typically employ a single discriminator operating on either spectral features or raw waveforms to evaluate signal realism and perceptual fidelity.
While \ac{mlp}-based discriminators~\citep{li2018speech} operate on spectral parameters such as \ac{lsf} and energy, \ac{cnn}-based architectures leverage a variety of input representations, including full-band \ac{lps}~\citep{eskimez2019speech}, \ac{mfcc}~\citep{li2019speech}, mel-spectrograms~\citep{haws2019cyclegan}, and raw waveforms~\citep{kim2019bandwidth,hao2020time}.
Subsequent works adopt multi-discriminator architectures to provide complementary adversarial guidance from multiple perceptual perspectives, enabling joint modeling of global structure and fine-grained temporal details. 
The majority of works~\citep{li2021real,su2021bandwidth,moliner2022behm,andreev2023hifiplusplus,gomez2023low,hauret2023eben,mandel2023aero,shuai2023mdctgan,soltanmohammadi2023low,yang2024sdnet,yu2024bae,kim2024audio,lu2024towards,lu2024multi,lee2025wave,yuan2025swinsrgan,zhao2025hifi,zhang2025vm,gao2026real,lee2026semambaplusplus} employ either the multi-scale discriminator (MSD) from MelGAN~\citep{kumar2019melgan} or the multi-period discriminator (MPD) from HiFi-GAN~\citep{kong2020hifi}. 
MSD evaluates waveforms at multiple temporal resolutions (full, $2\times$, and $4\times$ downsampled), allowing lower-resolution branches to capture \ac{lf} structures such as coarse spectral envelopes and prosodic patterns, while the full-resolution branch preserves \ac{hf} details and transient characteristics. 
In contrast, MPD analyzes periodically sampled waveform segments via \ac{2d} convolutions, explicitly modeling pitch-dependent harmonic regularity and periodic structure that conventional time-domain discriminators often fail to capture.

Alternative designs~\citep{su2021bandwidth,yoneyama2023nonparallel,hauret2023eben,lu2024towards,lu2024multi,zhao2025hifi,yuan2025swinsrgan, tamiti2025nldsi,lee2026semambaplusplus} further diversify discriminator architectures. 
\citet{su2021bandwidth} introduce a spectral-domain discriminator operating on log-mel spectrograms to enhance \ac{hf} reconstruction. 
\citet{yoneyama2023nonparallel} combine waveform-level~\citep{yamamoto2020parallel} and multi-band spectral discriminators~\citep{kumar2020nu} for independent frequency-band analysis. 
EBEN~\citep{hauret2023eben} employs a multi-scale ensemble discriminator on \ac{pqmf}-decomposed subbands with a full-scale MSD sub-discriminator to enforce cross-band coherence. 
\citet{lu2024towards,lu2024multi} introduce a multi-resolution amplitude discriminator (MRAD) and a multi-resolution phase discriminator (MRPD), inspired by the multi-resolution discriminator (MRD)~\citep{jang2021univnet}, operating across multiple \ac{stft} resolutions to alleviate spectral over-smoothing and phase distortion artifacts.
HiFi-SR~\citep{zhao2025hifi} addresses the limited phase sensitivity of MRD via a multi-band, multi-scale time-frequency discriminator (MBD) that jointly analyzes complex \ac{stft} components across time scales and frequency subbands. 
\citet{yuan2025swinsrgan} design a high-band multi-band discriminator (HB-MBD) that focuses on reconstructing \ac{hf} content in the \ac{mdct} domain. 
\citet{tamiti2025nldsi} introduce a multi-scale recurrence discriminator (MSRD) and a multi-resolution Lyapunov discriminator (MRLD), which impose nonlinear dynamical-system constraints to enable lightweight, depthwise-separable discriminator designs. 
In addition, SEMamba++~\citep{lee2026semambaplusplus} employs a multi-scale sub-band constant-Q transform (MS-SB-CQT) discriminator~\citep{gu2024multi} to better capture periodic and harmonic structures.

\subsubsection{Training Objective}
\ac{gan}-based \ac{bwe}/\ac{asr} methods typically optimize a generator $G$ that predicts a \ac{bb}/\ac{hr} estimate $\hat{\mathbf{y}} = G(\mathbf{x})$ from a \ac{bl}/\ac{lr} observation $\mathbf{x}$, together with discriminator(s) ${D}$ that distinguish $\hat{\mathbf{y}}$ from the ground truth $\mathbf{y}$. In practice, the generator objective is commonly formulated as a weighted combination of (i) an adversarial term that promotes perceptual realism, (ii) a reconstruction term that enforces signal fidelity, and (iii) an \ac{fm} term that stabilizes adversarial optimization.
The standard adversarial loss is defined as the min-max optimization by
\begin{equation}
\label{eq:gan_loss}
\min_{G} \max_{D} \;\; \mathbb{E}_{\mathbf{y}}\!\left[\log D(\mathbf{y})\right] + \mathbb{E}_{\mathbf{x}}\!\left[\log(1 - D(G(\mathbf{x})))\right],
\end{equation}
where \( D(\mathbf{y}) \) estimates the likelihood of the ground-truth \ac{bb}/\ac{hr} signal, and \( D(G(\mathbf{x})) \) measures the perceptual plausibility of the generated output. Through this dual-objective optimization, \ac{gan}-based approaches recover sharper spectral details and more realistic \ac{hf} contents compared to deterministic regression models.

The adversarial term drives distribution matching and perceptual realism, using binary \ac{ce} loss~\citep{li2018speech, eskimez2019speech, li2019speech, kim2019bandwidth, sautter2019artificial, haws2019cyclegan, hao2020time,shuai2023mdctgan,yoneyama2023nonparallel,gao2026harmonic}, hinge loss~\citep{li2021real,hauret2023eben,mandel2023aero,soltanmohammadi2023low,gomez2023low,yang2024sdnet,lu2024towards,lu2024multi,bao2025frequency}, or \ac{lsgan} loss~\citep{andreev2023hifiplusplus,kim2024audio,yu2024bae,zhao2025hifi,yuan2025swinsrgan,zhang2025vm,dai2025sfnet,liu2025hwb,gao2026real,lee2026semambaplusplus}.

The binary \ac{ce} loss is expressed as
\begin{equation}
\label{eq:ce_g}
\mathcal{L}_G^{\text{CE}} = 
-\mathbb{E}_{\hat{\mathbf{y}}}[\log D(\hat{\mathbf{y}})],
\end{equation}
\begin{equation}
\label{eq:ce_d}
\mathcal{L}_D^{\text{CE}} = 
-\mathbb{E}_{\mathbf{y}}[\log D(\mathbf{y})] - 
\mathbb{E}_{\hat{\mathbf{y}}}[\log(1 - D(\hat{\mathbf{y}}))].
\end{equation}
The hinge loss is formulated as
\begin{equation}
\label{eq:hinge_g}
\mathcal{L}_G^{\text{hinge}} = 
-\mathbb{E}_{\hat{\mathbf{y}}}[D(\hat{\mathbf{y}})],
\end{equation}
\begin{equation}
\label{eq:hinge_d}
\mathcal{L}_D^{\text{hinge}} = 
\mathbb{E}_{\hat{\mathbf{y}}}[\max(0, 1 + D(\hat{\mathbf{y}}))] + 
\mathbb{E}_{\mathbf{y}}[\max(0, 1 - D(\mathbf{y}))].
\end{equation}
The \ac{lsgan} loss is defined as
\begin{equation}
\label{eq:lsgan_g}
\mathcal{L}_G^{\text{LS}} = 
\mathbb{E}_{\hat{\mathbf{y}}}[(D(\hat{\mathbf{y}}) - 1)^2],
\end{equation}
\begin{equation}
\label{eq:lsgan_d}
\mathcal{L}_D^{\text{LS}} = 
\mathbb{E}_{\mathbf{y}}[(D(\mathbf{y}) - 1)^2] + 
\mathbb{E}_{\hat{\mathbf{y}}}[D(\hat{\mathbf{y}})^2].
\end{equation}
Early \ac{gan}-based designs often adopt the \ac{ce} loss formulation, whereas later works favor hinge loss or \ac{lsgan} objectives for improved gradient behavior and training stability.

To enforce fidelity to the target signal and preserve spectral structure, \ac{gan}-based \ac{bwe}/\ac{asr} methods augment adversarial training with time-domain and/or frequency-domain reconstruction losses. A widely adopted choice is the \ac{mrstft} loss~\citep{yamamoto2020parallel}, which combines spectral convergence and log-magnitude terms across multiple \ac{stft} configurations (e.g., varying FFT sizes, window lengths, and hop sizes):
\begin{equation}
\label{eq:mrstft}
\mathcal{L}_{G}^{\text{STFT}} 
= \frac{1}{M} \sum_{m=1}^{M} 
\mathbb{E}_{\mathbf{y},\hat{\mathbf{y}}} 
\left[
\mathcal{L}_{\text{sc}}^{m}(\mathbf{y},\hat{\mathbf{y}}) 
+ 
\mathcal{L}_{\text{mag}}^{m}(\mathbf{y},\hat{\mathbf{y}})
\right],
\end{equation}
where $M$ is the number of \ac{stft} configurations, $G$ denotes the generator.
The spectral convergence loss $\mathcal{L}_{\text{sc}}$ and log-magnitude loss $\mathcal{L}_{\text{mag}}$ are defined as
\begin{equation}
\label{eq:lsc}
\mathcal{L}_{\text{sc}}(\mathbf{y},\hat{\mathbf{y}}) 
= 
\frac{ \left\|\, | \text{STFT}(\mathbf{y}) | - | \text{STFT}(\hat{\mathbf{y}}) | \,\right\|_{F} }
     { \left\|\, | \text{STFT}(\mathbf{y}) | \,\right\|_{F} },
\end{equation}
\begin{equation}
\label{eq:lmag}
\mathcal{L}_{\text{mag}}(\mathbf{y},\hat{\mathbf{y}}) 
= 
\left\|
\log | \text{STFT}(\mathbf{y}) |
-
\log | \text{STFT}(\hat{\mathbf{y}}) |
\right\|_{1},
\end{equation}
where $|\text{STFT}(\cdot)|$ denotes the magnitude of STFT, and $\|\cdot\|_{F}$ and $\|\cdot\|_{1}$ denote the Frobenius norm and $l_1$ norm, respectively. 

Another widely adopted objective is the mel-spectrogram \ac{mae} loss, which evaluates reconstruction accuracy in a perceptually weighted mel-frequency domain. Its formulation is given by
\begin{equation}
\label{eq:mel_loss}
\mathcal{L}_{G}^{\text{mel}} = \left\| \text{Mel}(\mathbf{y}) - \text{Mel}(\hat{\mathbf{y}}) \right\|_{1},
\end{equation}
where $\text{Mel}(\cdot)$ denotes the mel-spectrogram transformation. Both objectives act as perceptually motivated reconstruction losses, reflecting the nonlinear frequency sensitivity of human hearing and encouraging perceptually plausible spectral detail in the reconstructed signals.

To stabilize adversarial optimization and suppress unnatural
artifacts, the \ac{fm} loss aligns intermediate discriminator features between real and generated samples:
\begin{equation}
\label{eq:fmloss}
\mathcal{L}_G^{\text{FM}} 
= 
\mathbb{E}_{\mathbf{y},\hat{\mathbf{y}}}
\left[
\sum_{i=1}^{K}
\frac{1}{N_i}
\left\|
D_i(\mathbf{y}) - D_i(\hat{\mathbf{y}})
\right\|_{1}
\right],
\end{equation}
where $K$ denotes the number of discriminator layers, $D_i(\cdot)$ represents the intermediate feature map at the $i$-th layer, and $N_i$ denotes the number of elements in that feature map. These components are typically combined into a weighted generator objective:
\begin{equation}
\label{eq:total_g_loss}
\mathcal{L}_G = 
\lambda_{\text{adv}} \mathcal{L}_{G}^{\text{adv}} 
+ \lambda_{\text{rec}} \mathcal{L}_{G}^{\text{rec}} 
+ \lambda_{\text{FM}} \mathcal{L}_{G}^{\text{FM}},
\end{equation}
where $\mathcal{L}_{G}^{\text{adv}}$, $\mathcal{L}_{G}^{\text{rec}}$, and $\mathcal{L}_{G}^{\text{FM}}$ denote the adversarial, reconstruction, and \ac{fm} losses, respectively, and $\lambda_{\text{adv}}, \lambda_{\text{rec}}, \lambda_{\text{FM}}$ are weighting coefficients that balance perceptual naturalness and spectral fidelity. Simultaneously, the discriminator is optimized to minimize $\mathcal{L}_{D}$. Moreover, some works incorporate task-aware perceptual constraints, such as matching embeddings from a pretrained automatic speech recognition network~\citep{li2019speech}, to enhance downstream performance.

\subsection{Diffusion Model}
\begin{figure}[t]
\centering
\begin{subfigure}[t]{0.495\linewidth}
\centering
\includegraphics[width=\linewidth]{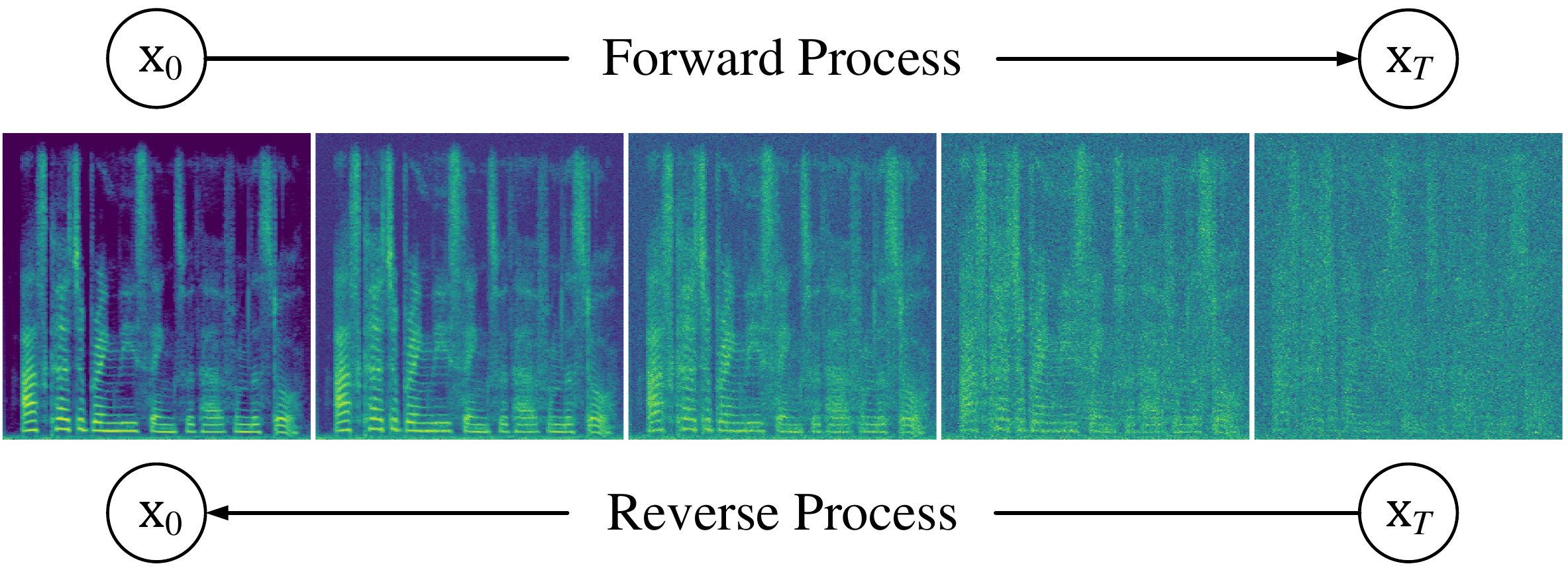}
\caption{Diffusion}
\label{fig_first_case}
\end{subfigure}\hfill
\begin{subfigure}[t]{0.495\linewidth}
\centering
\includegraphics[width=\linewidth]{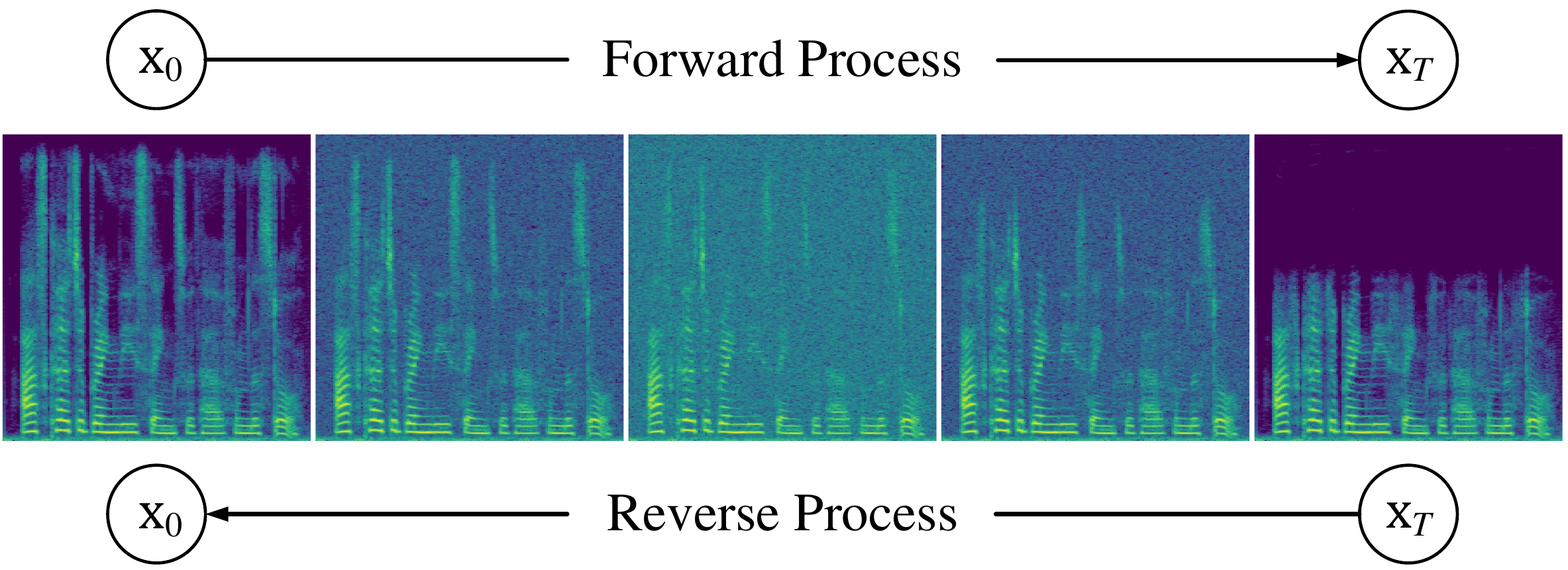}
\caption{Bridge}
\label{fig_second_case}
\end{subfigure}
\caption{\textbf{Illustration of diffusion and bridge processes on audio spectrograms.} \textbf{(a)} Diffusion: the forward process progressively corrupts the clean audio spectrogram $\mathbf{x}_{0}$ by injecting noise, whereas the learned reverse process iteratively denoises a heavily perturbed sample $\mathbf{x}_{T}$ to recover $\mathbf{x}_{0}$. \textbf{(b)} Bridge: the forward bridge process degrades the \ac{hr} spectrogram $\mathbf{x}_{0}$ into a \ac{lr} spectrogram $\mathbf{x}_{T}$ through progressive bandwidth reduction with stochastic perturbations, while the reverse bridge process reconstructs $\mathbf{x}_{0}$ from $\mathbf{x}_{T}$.}
\label{fig:fig_diff_and_bridge}
\end{figure}

Diffusion models~\citep{ho2020denoising,rombach2021highresolution,ho2022classifierfree,gao2023masked,li2024pruning,lu2025hdc,song2026diffusion}, also known as diffusion-based generative models or \acp{sgm}, constitute a class of generative models that synthesize data by learning to invert a progressively applied Gaussian noising process. A diffusion model comprises two coupled stochastic processes: a forward diffusion process and a reverse sampling process. In the forward process, progressively increasing Gaussian noise is added to a clean signal $\mathbf{x}_0$ according to
\begin{equation}
\mathbf{x}_t = \sqrt{\alpha_t}\, \mathbf{x}_{t-1} + \sqrt{1-\alpha_t}\,\boldsymbol{\epsilon},
\end{equation}
\begin{equation}
\boldsymbol{\epsilon} \sim \mathcal{N}(\mathbf{0}, \displaystyle \mI), \quad t = 1,\ldots,T,
\end{equation}
where $\mathbf{x}_t$ denotes the intermediate signal at the diffusion step $t$, $\alpha_t = 1 - \beta_t$, and $\{\beta_t\}_{t=1}^{T}$ denotes a predefined noise variance schedule. This forward process progressively transforms the data distribution into an isotropic Gaussian distribution. The reverse process is parameterized by a neural network $\epsilon_{\theta}(\mathbf{x}_t, t)$ that predicts the injected noise, or equivalently the denoising direction, at each diffusion step $t$, and generates samples via iterative denoising according to
\begin{equation}
\mathbf{x}_{t-1}
= \frac{1}{\sqrt{\alpha_t}}
\left(
\mathbf{x}_t - \frac{1 - \alpha_t}{\sqrt{1 - \bar{\alpha}_t}}\,
\epsilon_{\theta}(\mathbf{x}_t, t)
\right)
+ \sigma_t \mathbf{z},
\end{equation}
\begin{equation}
\mathbf{z} \sim \mathcal{N}(\mathbf{0}, \displaystyle \mI), \quad t = T,\ldots,1,
\end{equation}
with $\bar{\alpha}_t = \prod_{i=1}^{t} \alpha_i$ and $\sigma_t = \sqrt{\tfrac{1 - \bar{\alpha}_{t-1}}{1 - \bar{\alpha}_t}\, \beta_t}$. By learning to estimate the denoising direction at each step accurately, the reverse process enables the generation of samples following the underlying training data distribution. The forward and reverse diffusion processes are illustrated in Fig.~\ref{fig:fig_diff_and_bridge}~(\subref{fig_first_case}). Owing to their relatively stable training, flexible distribution modeling, and demonstrated perceptual fidelity, diffusion models have gained increasing attention across a wide range of audio generation and restoration tasks.

NU-Wave~\citep{lee2021nu} introduces waveform-to-waveform conditional diffusion for \ac{ssr}, adopting a vocoder-inspired residual \ac{bidilconv} backbone from DiffWave~\citep{kong2020diffwave} and injecting the \ac{lr} waveform via a redesigned conditioner with an enlarged receptive field to enable accurate upsampling from $16$ kHz or $24$ kHz to $48$ kHz. NU-Wave2~\citep{han2022nu} further improves harmonic reconstruction by introducing \ac{stfc} to provide stronger global spectral context and \ac{bsft} to explicitly condition on the retained bandwidth, enabling a single model to handle diverse input sampling rates with fewer parameters than NU-Wave~\citep{lee2021nu}. FastWave~\citep{kuznetsov2026fastwave} further lightens NU-Wave2~\citep{han2022nu} with architectural modifications to reduce the computational complexity for any-to-$48$ kHz \ac{asr}.

To mitigate computational costs of waveform-level diffusion, AudioSR~\citep{liu2024audiosr} performs \ac{asr} in a compact latent space obtained via a \ac{vae} encoder. Inspired by AudioLDM~\citep{liu2023audioldm}, the \ac{ldm} predicts \ac{hr} mel-spectrograms from their \ac{lr} counterparts, followed by waveform synthesis using a HiFi-GAN vocoder~\citep{kong2020hifi}. Building on this framework, FlashSR~\citep{im2025flashsr} distills the multi-step \ac{ldm} into a single-step model via Flash Diffusion~\citep{chadebec2025flash} and distribution-matching objectives, enabling efficient sampling while preserving perceptual quality. It further integrates a BigVGAN-based vocoder~\citep{lee2022bigvgan}, jointly conditioned on the predicted mel-spectrogram and input \ac{lr} waveform, eliminating explicit \ac{lf} refinement and enhancing \ac{hf} reconstruction. \citet{jin2026inference} propose an inference-time scaling diffusion (IS-Diffusion) \ac{asr} framework built on AudioSR~\citep{liu2024audiosr}, which generates multiple \ac{hr} candidates and leverages task-specific verifiers with zero-order search to select optimal samples, thereby reducing stochastic variance and improving spectral fidelity.

UDM+~\citep{yu2023conditioning} reframes \ac{ssr} as a frequency-domain inpainting problem and proposes a reverse-sampling strategy that propagates \ac{lr} information through the sampling trajectory, rather than relying solely on auxiliary conditioning of the denoiser. Combined with an \ac{udm}, this formulation improves robustness across varying upsampling ratios and downsampling filters. 
SWiBE~\citep{lin2024swibe} targets noise-robust \ac{bwe} by parameterizing a stochastic diffusion trajectory that progressively expands bandwidth in the complex spectrogram domain, using a U-Net-based \ac{sgm} (NCSN++~\citep{song2020score}) to estimate score functions along the trajectory. Vector Quantized Diffusion (VQ-Diffusion)~\citep{fang2025vector} performs \ac{bwe} directly in the discrete token space derived from a \ac{nac}, mitigating \ac{hf} loss associated with mel-spectrogram front-ends. It combines Descript Audio Codec (DAC)-based tokenization, a discrete denoising diffusion model (D3PM) for token reconstruction, and a Mamba-2 backbone to capture long-range temporal dependencies for high-fidelity \ac{bwe}.

\citet{lemercier2023analysing} extend diffusion-based generative modeling to speech restoration in the complex time-frequency domain, comparing it with a discriminative counterpart using the same \ac{dnn} architecture. This method employs iterative denoising to capture complex degradations and shows consistent improvements over discriminative models, particularly for \ac{bwe} and dereverberation.
UNIVERSE~\citep{serra2022universal} proposes a unified generative framework for universal \ac{se} by decoupling restoration into a multi-resolution conditioning network and a score-based diffusion generator, where mixture density networks (MDNs) provide robust spectral guidance. Building on this, UNIVERSE++~\citep{scheibler2024universal} enhances stability and fidelity via anti-aliasing, HiFi-GAN-based adversarial loss, and low-rank adaptation (LoRA)~\citep{hu2022lora} with phoneme-level constraints to reduce hallucinations and improve linguistic consistency.

Beyond supervised conditional modeling, diffusion priors can be leveraged for inverse problems with uncertain or unknown degradations. BABE~\citep{moliner2024blind} addresses zero-shot blind \ac{bwe} of historical recordings by combining a pretrained unconditional diffusion prior with generalized diffusion posterior sampling, jointly estimating a parametric low-pass degradation operator during iterative \ac{hf} content reconstruction.

\subsection{Flow-based Model}
\label{sec:flow_model}

Flow-based generative models~\citep{ho2019flowplusplus,lipman2023flowmatch,tong2023improving,gat2024discrete,zhou2026emotional} learn transport mappings that transform a simple source distribution into the conditional distribution of \ac{bb}/\ac{hr} audio given a \ac{bl}/\ac{lr} observation. In the current \ac{bwe}/\ac{asr} literature, this family can be broadly divided into two branches: (i) Glow-style discrete normalizing flows, which optimize exact conditional likelihood via a finite sequence of invertible transformations, and (ii) continuous normalizing flows trained with flow matching, which learn a continuous transport governed by an \ac{ode} by regressing a target velocity field~\citep{pooladian2023}. Although both branches are flow-based, they occupy distinct design regimes. Glow-style models emphasize exact likelihood computation and one-pass invertibility. In contrast, flow matching provides greater flexibility in transport path design and conditioning mechanisms, but requires solving a continuous-time \ac{ode} during sampling. Their relationship and trade-offs are summarized in Table~\ref{tab:glow_vs_fm}.

\begin{table}[t]
\centering
\renewcommand{\arraystretch}{1.25}
\setlength{\tabcolsep}{3pt}
\caption{\textbf{Key distinctions between Glow-style discrete normalizing flows and flow matching.} Glow-style models use constrained invertible architectures for exact likelihoods, while flow matching adopts simulation-free vector-field learning for flexible probability paths with iterative inference.}
\vspace{0pt}
\label{tab:glow_vs_fm}

\small
\begin{tabular*}{\linewidth}{@{}c@{\hspace{25pt}}c@{\hspace{25pt}}c@{\hspace{25pt}}c@{}}
\toprule
\textbf{Modeling} & \textbf{Mapping} & \textbf{Training} & \textbf{Inference} \\
\midrule

Glow-style Flow & Invertible Transformation & Exact Likelihood Maximization & Single Forward Pass \\
\addlinespace[5pt]

Flow Matching & Continuous-time \ac{ode} & Velocity Field Regression & One or Multi-step \ac{ode} \\

\bottomrule
\end{tabular*}
\end{table}

A Glow-style conditional normalizing flow defines an invertible mapping
\begin{equation}
\mathbf{y} = f_{\theta}(\mathbf{z}\mid\mathbf{c}),
\end{equation}
where $\mathbf{y}$ denotes the target \ac{bb}/\ac{hr} audio, $\mathbf{z}$ is a latent variable drawn from a simple base distribution $p(\mathbf{z})$, which is typically chosen as $\mathcal{N}(\mathbf{0}, \mathbf{I})$, and $\mathbf{c}$ represents the conditioning information extracted from the \ac{bl}/\ac{lr} observation. The corresponding exact conditional likelihood is obtained via the change-of-variables formula given as
\begin{equation}
\log p_{\theta}(\mathbf{y}\mid\mathbf{c})
=
\log p(\mathbf{z})
+
\log\left|
\det
\frac{\partial f_{\theta}^{-1}(\mathbf{y}\mid\mathbf{c})}{\partial \mathbf{y}}
\right|,
\end{equation}
where $\det(\cdot)$ denotes the determinant of the Jacobian matrix associated with the inverse mapping. WSRGlow~\citep{zhang2021wsrglow} serves as a representative Glow-style model for \ac{asr}, integrating invertible flow-based transformations with WaveNet-style affine coupling to maximize the exact conditional likelihood of the \ac{hr} waveforms given the \ac{lr} observations. Its dual-conditioning design combines a time-domain \ac{lr} encoder with an \ac{stft}-based encoder, enabling the fusion of complementary temporal and spectral cues for enhanced \ac{hf} reconstruction.

More recent research has increasingly explored flow matching, which trains a continuous normalizing flow by learning a conditional velocity field along a prescribed probability path~\citep{lipman2023flowmatch,pooladian2023,yun2025flowhigh,ku2025generative}. The transport dynamics are controlled by
\begin{equation} \frac{d}{dt}\mathbf{y}_t = v_{\theta}(\mathbf{y}_t, t \mid \mathbf{c}), \end{equation}
where $\mathbf{y}_t \sim p_t(\cdot \mid \mathbf{c})$ denotes the intermediate state at time $t$, $v_{\theta}$ is the learned conditional velocity field, and $\mathbf{c}$ represents the condition derived from the degraded input. The process evolves from a source distribution $p_0(\cdot \mid \mathbf{c})$ to the target conditional \ac{hr} audio distribution $p_1(\cdot \mid \mathbf{c})$ at $t=1$. In contrast to Glow-style flows, which optimize exact likelihood, recent \ac{bwe}/\ac{asr} systems adopt flow matching to learn this transport via conditional velocity regression.

Within this branch, FLowHigh~\citep{yun2025flowhigh} demonstrates efficient single-step sampling for \ac{ssr} through carefully designed conditional transport paths. Specifically, it introduces a data-dependent source prior and a post-processing step that restores the observed \ac{lf} band, enabling high-quality single-step sampling. \citet{hernandez2026single} extend this paradigm to controllable music \ac{bwe} by incorporating Dynamic Spectral Contours and classifier-free guidance, enabling finer control over the reconstructed bandwidth content.
A complementary direction scales flow matching through enhanced representations and training strategies. \citet{ku2025generative} pretrain a generative speech foundation model (GSFM) directly on complex-valued \ac{stft} coefficients, eliminating the need for a separate vocoder and supporting multiple restoration tasks, including \ac{bwe}. Along a related vocoder-free direction, UniverSR~\citep{choi2025universr} models the conditional distribution of complex spectral coefficients via flowmatching and reconstructs audio using \ac{istft}, yielding a unified \ac{asr} framework without a dedicated neural vocoder.

Recent work further expands conditioning mechanisms and improves computational efficiency. SAGA-SR~\citep{im2026saga} augments flow matching with semantic guidance from text and acoustic cues derived from spectral roll-off, employing a diffusion-Transformer backbone to enable any-rate upsampling to $44.1$ kHz. CodecFlow~\citep{zhang2026codecflow}, in contrast, performs \ac{bwe} in a neural codec latent space, where a voicing-aware conditional flow converter and a structure-constrained \ac{rvq} module mitigate latent mismatch and improve the compute-quality trade-off. To facilitate real-time deployment, Stream FM~\citep{welker2025real} and Real-time Flow~\citep{hsieh2026towards} adopt few-step sampling strategies within their flow matching formulations, effectively bridging the gap between generative quality and low-latency requirements. LatentFlowSR~\citep{liu2026latentflowsr} combines conditional flow matching with a noise-robust autoencoder in a continuous latent space, enabling efficient one-step ODE sampling for reconstructing missing \ac{hf} content across speech, music, and sound effects.


Overall, recent flow-based \ac{bwe}/\ac{asr} research has increasingly emphasized flow matching over Glow-style exact-likelihood models, favoring flexible conditional transport and efficient sampling strategies. Nevertheless, the computational advantages of flow matching depend on the chosen probability paths, model architectures, and numerical solvers.

\subsection{Bridge Model}
Bridge models, often described as Schr\"odinger bridges in recent audio machine learning literature~\citep{chen2023schrodinger,wang2024framebridge,li2025bridge, li2025audio, shihaudio}, are closely related to diffusion and score-based models but instead learn a stochastic transport between endpoint marginals.
Diffusion models start from an approximately Gaussian endpoint induced by progressive noising, whereas bridge models directly connect prescribed source and target distributions, as illustrated in Fig.~\ref{fig:fig_diff_and_bridge}.
However, unlike the classical Schr\"odinger bridge~\citep{leonard2013survey,de2024schrodinger}—an unsupervised optimal transport formulation solved via iterative proportional fitting on unpaired distributions—most current \ac{bwe}/\ac{asr} approaches~\citep{li2025bridge, li2025audio, shihaudio} adopt supervised training with paired data and single-pass regression, making them more closely related to \textit{bridge matching}~\citep{shi2023diffusion,zhou2023denoising,peluchetti2023diffusion}.
This formulation is particularly appealing for \ac{bwe}/\ac{asr} because the observed \ac{bl}/\ac{lr} signal itself provides an informative starting point for reconstructing the \ac{bb}/\ac{hr} target.

Let $p_{\mathrm{src}}$ and $p_{\mathrm{tgt}}$ denote the source and target distributions, and let $\{\mathbf{z}_t\}_{t\in[0,T]}$ denote the corresponding transport trajectory. For \ac{bwe}/\ac{asr}, $p_{\mathrm{src}}$ is induced by degraded \ac{bl}/\ac{lr} observations, while $p_{\mathrm{tgt}}$ corresponds to the \ac{bb}/\ac{hr} data distribution.
Starting from a reference diffusion process,
\begin{equation}
d\mathbf{z}_t = f(\mathbf{z}_t,t)\,dt + g(t)\,d\mathbf{w}_t, \qquad t \in [0,T],
\end{equation}
the Schr\"odinger bridge problem seeks a path measure $p \in \mathcal{P}[0,T]$ that matches the endpoint marginals while remaining close to a reference measure $p_{\mathrm{ref}}$:
\begin{equation}
\min_{p \in \mathcal{P}[0,T]} D_{\mathrm{KL}}(p \,\|\, p_{\mathrm{ref}})
\quad \text{s.t.} \quad
p_0 = p_{\mathrm{src}}, \; p_T = p_{\mathrm{tgt}}.
\end{equation}
This problem admits an equivalent characterization via paired forward and backward \acp{sde} with corrected drifts:
\begin{equation}
d\mathbf{z}_t
=
\left[
f(\mathbf{z}_t,t)
+ g^2(t)\nabla_{\mathbf{z}} \log \Psi_t(\mathbf{z}_t)
\right] dt
+ g(t)\, d\mathbf{w}_t,
\label{eq:bridge_forward}
\end{equation}
\begin{equation}
d\mathbf{z}_t
=
\left[
f(\mathbf{z}_t,t)
- g^2(t)\nabla_{\mathbf{z}} \log \hat{\Psi}_t(\mathbf{z}_t)
\right] dt
+ g(t)\, d\bar{\mathbf{w}}_t,
\label{eq:bridge_backward}
\end{equation}
where $\Psi_t$ and $\hat{\Psi}_t$ are time-dependent Schr\"odinger potentials. Compared with conventional diffusion models initialized from Gaussian noise, bridge-based formulations can exploit informative degraded observations as source distributions, potentially enabling more efficient reconstruction trajectories depending on the bridge formulation and sampling strategy.

Bridge-SR~\citep{li2025bridge} is an early bridge-based framework for \ac{asr}. Operating directly in the waveform domain, it uses the observed \ac{lr} waveform as an informative prior for the \ac{hr} target and learns the corresponding score functions for efficient data-to-data generation. The study reports favorable quality-efficiency trade-offs under its evaluated settings using a lightweight backbone and few-step synthesis. AudioLBM~\citep{li2025audio} extends this idea from waveform-space transport to continuous latent-space transport. By compressing audio into a continuous latent representation and performing latent-to-latent bridging, it better aligns the generative process with \ac{lr}-to-\ac{hr} upsampling. It further introduces frequency-aware conditioning on the source and target bands to enable any-to-any upsampling during training, while cascaded latent bridge models and prior augmentation support \ac{asr} beyond $48$~kHz, including $96$ and $192$~kHz, across speech, music, and general sound effects. 
A2SB~\citep{shihaudio} formulates \ac{bwe} and audio inpainting as a unified spectrogram inpainting problem within the Schr\"odinger bridge framework, enabling end-to-end waveform generation without relying on a separate vocoder. The model further incorporates a factorized magnitude-phase representation and sliding-window inference to support \ac{hr} audio restoration at $44.1$ kHz over long music segments.

%% file: sec/7_quantitative_final.tex
\section{Quantitative Analysis}
\label{sec:quan}

To enable a fair and direct comparison across different \ac{asr}/\ac{bwe} paradigms, we re-evaluate representative discriminative, GAN-based, diffusion-based, and flow-based methods under a unified experimental setup. This quantitative analysis provides stronger empirical evidence for the trade-offs among reconstruction accuracy, perceptual quality, model complexity, and computational efficiency.

\subsection{Experimental Setup}

We conduct unified experiments on the VCTK corpus~\citep{yamagishi2012english}, which consists of 108 speakers, for two \ac{asr} tasks: 8 kHz $\rightarrow$ 16 kHz and 16 kHz $\rightarrow$ 48 kHz. The original 48 kHz speech recordings are first low-pass filtered and downsampled to 16 kHz, followed by the same procedure to obtain 8 kHz signals. Both stages employ Chebyshev Type-I filtering and polyphase resampling implemented in \texttt{SciPy}, with zero-phase filtering, randomly sampled filter orders (1--11) and passband ripples ($10^{-9}$, $10^{-6}$, $10^{-3}$, 1, and 5 dB) during training, and a fixed eighth-order filter with 0.05 dB ripple for validation and testing. Both target and degraded waveforms are independently normalized to a root mean square (RMS) amplitude of 0.05. We adopt speaker-disjoint training, validation, and test splits containing 90, 10, and 8 speakers, respectively.

We evaluate ten representative methods spanning four major learning paradigms to systematically compare their reconstruction quality and computational efficiency. The discriminative models include AudioUNet~\citep{kuleshov2017audio}, TFiLM~\citep{birnbaum2019temporal}, and AFiLM~\citep{rakotonirina2021self}; the \ac{gan}-based models include AERO~\citep{mandel2023aero}, AP-BWE~\citep{lu2024towards}, and VM-ASR~\citep{zhang2025vm}; the diffusion-based models include NU-Wave2~\citep{han2022nu} and UDM+~\citep{yu2023conditioning}; and the flow-based models include FLowHigh~\citep{yun2025flowhigh} and UniverSR~\citep{choi2025universr}. All methods are trained and evaluated using their official implementations and recommended configurations. To ensure consistent comparisons, we adopt identical VCTK speaker splits, degradation procedures, and evaluation protocols while preserving each method's original architecture and training objectives.

All methods are evaluated on the same test set using LSD metrics (LSD, LSD-LF, and LSD-HF) to assess overall, \ac{lf}, and \ac{hf} spectral reconstruction fidelity, respectively, and ViSQOL to evaluate perceptual audio quality. We additionally report model size and real-time factor (RTF) to assess computational efficiency. Parameter counts include all neural components required for end-to-end waveform inference, while RTF is defined as the ratio of total end-to-end inference time to generated audio duration, with lower values indicating faster inference. All models are trained and evaluated on a single NVIDIA RTX A6000 GPU.

\begin{table*}[t]
\centering
\captionsetup{skip=4pt}
\caption{\textbf{Quantitative comparison of representative SR methods on VCTK for the 8 kHz $\rightarrow$ 16 kHz and 16 kHz $\rightarrow$ 48 kHz tasks.} Lower values indicate better performance for LSD, LSD-LF, and LSD-HF, whereas higher values indicate better perceptual quality for ViSQOL.}
\label{tab:main_results}

\begingroup
\setlength{\tabcolsep}{3.5pt}
\renewcommand{\arraystretch}{1.30}

\resizebox{\textwidth}{!}{%
\begin{tabular}{llc|cccc|cccc}
\toprule
\multirow[c]{2}{*}[-0.4ex]{\textbf{Model Family}}
& \multirow[c]{2}{*}[-0.4ex]{\textbf{Method}}
& \multirow[c]{2}{*}[-0.4ex]{\textbf{\# Params}}
& \multicolumn{4}{c|}{\textbf{8~kHz $\rightarrow$ 16~kHz}}
& \multicolumn{4}{c}{\textbf{16~kHz $\rightarrow$ 48~kHz}} \\
\cmidrule(lr){4-7}
\cmidrule(lr){8-11}
&
&
& \textbf{LSD $\downarrow$}
& \textbf{LSD-LF $\downarrow$}
& \textbf{LSD-HF $\downarrow$}
& \textbf{ViSQOL $\uparrow$}
& \textbf{LSD $\downarrow$}
& \textbf{LSD-LF $\downarrow$}
& \textbf{LSD-HF $\downarrow$}
& \textbf{ViSQOL $\uparrow$} \\
\midrule

--
& Input
& --
& 2.227 & 0.217 & 3.136 & 4.199
& 2.976 & 0.220 & 3.642 & 1.821 \\

\midrule

\multirow[c]{3}{*}{Discriminative}
& AudioUNet
& 70.97M
& 1.563 & 0.193 & 2.195 & 4.288
& 2.486 & 0.185 & 3.042 & 1.892 \\

& TFiLM
& 68.23M
& 1.538 & 0.169 & 2.167 & 4.296
& 2.367 & 0.172 & 2.897 & 1.936 \\

& AFiLM
& 134.72M
& 1.479 & 0.220 & 2.078 & 4.310
& 2.195 & 0.201 & 2.685 & 2.075 \\

\midrule

\multirow[c]{3}{*}{GAN}
& AERO
& 36.36M
& 1.094 & 0.409 & 1.472 & 4.435
& 1.046 & 0.675 & 1.192 & 2.837 \\

& AP-BWE
& 29.76M
& 0.831 & 0.186 & 1.157 & 4.649
& 0.749 & 0.228 & 0.902 & 3.367 \\

& VM-ASR
& 3.01M
& 1.177 & 0.608 & 1.545 & 4.392
& 0.997 & 0.671 & 1.120 & 2.625 \\

\midrule

\multirow[c]{2}{*}{Diffusion}
& NU-Wave2
& 1.71M
& 1.453 & 0.316 & 2.003 & 4.359
& 1.173 & 0.528 & 1.386 & 2.653 \\

& UDM+
& 6.26M
& 1.023 & 0.250 & 1.420 & 4.405
& 0.940 & 0.485 & 1.096 & 2.741 \\


\midrule

\multirow[c]{2}{*}{Flow}
& FLowHigh
& 48.84M
& 0.781 & 0.219 & 1.078 & 4.652
& 0.901 & 0.223 & 1.091 & 3.556 \\

& UniverSR
& 57.03M
& 0.841 & 0.285 & 1.151 & 4.533
& 0.967 & 0.258 & 1.169 & 3.128 \\

\bottomrule
\end{tabular}%
}

\endgroup
\vspace{-1.0mm}
\end{table*}

\subsection{Results}
Table~\ref{tab:main_results} presents a unified quantitative comparison of representative discriminative, GAN-based, diffusion-based, and flow-based methods on the 8 kHz $\rightarrow$ 16 kHz and 16 kHz $\rightarrow$ 48 kHz tasks. Overall, generative methods substantially outperform discriminative models in \ac{hf} reconstruction and perceptual quality. Among these approaches, AP-BWE and FLowHigh achieve the strongest performance. AP-BWE obtains the lowest LSD and LSD-HF for 16 kHz $\rightarrow$ 48 kHz, whereas FLowHigh achieves the highest ViSQOL in both tasks and the lowest LSD for 8 kHz $\rightarrow$ 16 kHz. UniverSR also performs consistently well across both settings. These results indicate that flow-based methods provide a favorable balance between spectral fidelity and perceptual quality, while carefully designed GAN-based methods remain highly competitive in spectral reconstruction. The comparisons further show that model size is not a reliable predictor of performance, as several compact generative models outperform larger discriminative counterparts (e.g., UDM+ vs. TFiLM). This suggests that architectural design and learning objectives are more influential than parameter count alone.

\begin{figure*}[t]
\centering
\begin{subfigure}[t]{0.495\textwidth}
\centering
\includegraphics[width=\linewidth]{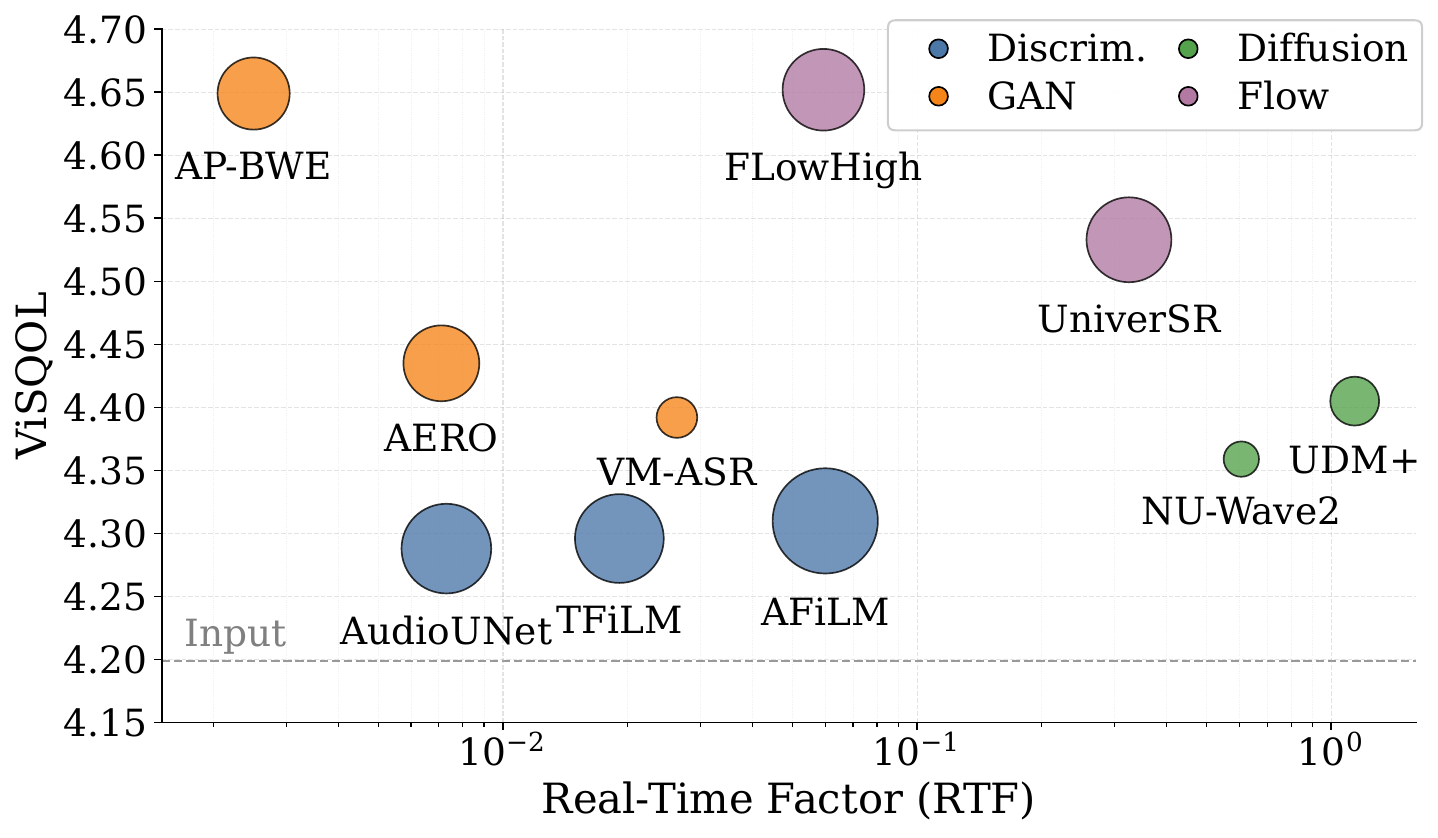}
\caption{}
\label{fig:rtf_visqol_bubble}
\end{subfigure}
\hfill
\begin{subfigure}[t]{0.495\textwidth}
\centering
\includegraphics[width=\linewidth]{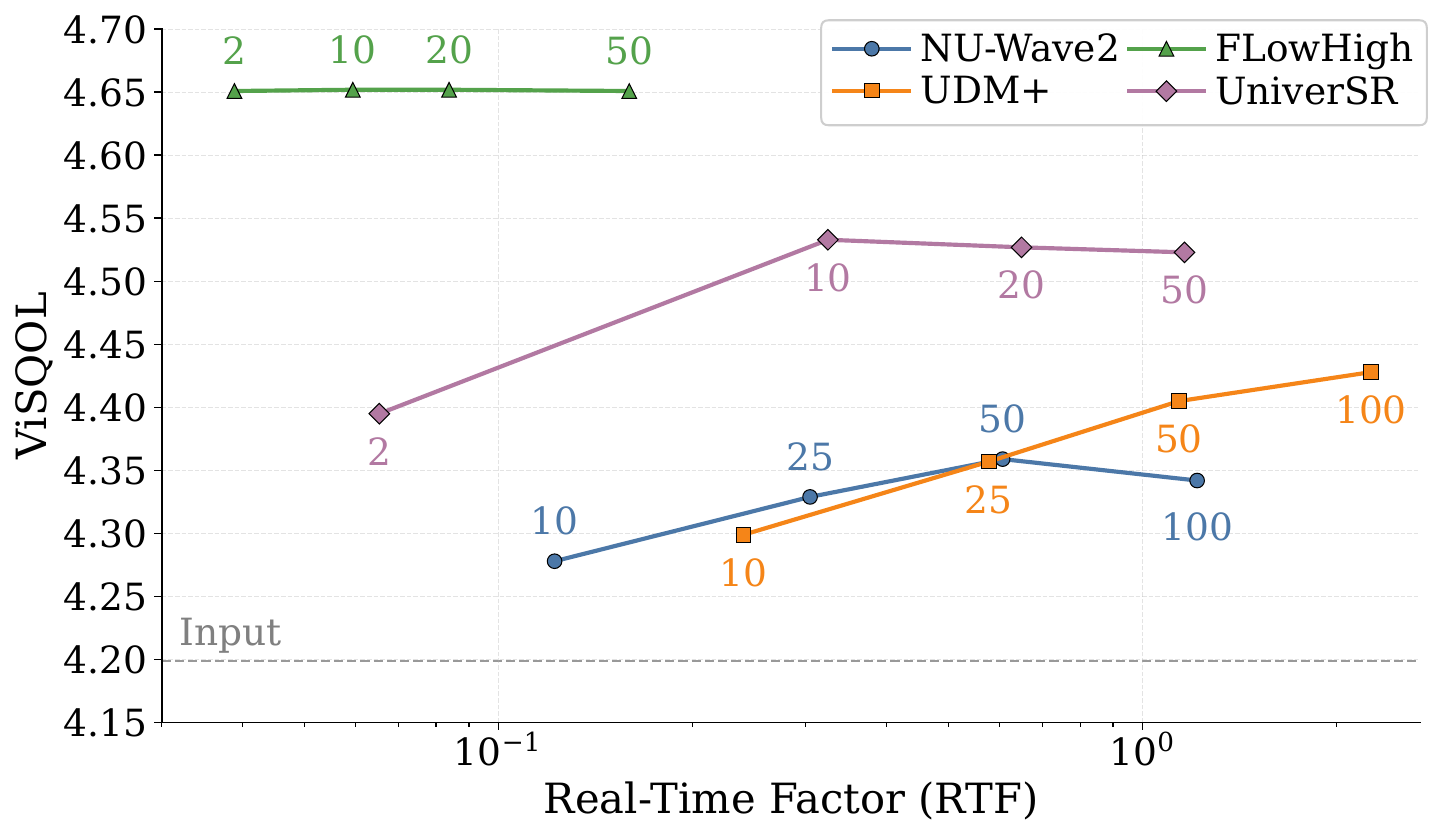}
\caption{}
\label{fig:rtf_visqol_nfe}
\end{subfigure}
\caption{\textbf{Trade-off between perceptual quality and computational efficiency for 8 kHz $\rightarrow$ 16 kHz SR on VCTK.} \textbf{(a)} Comparison of model size, inference speed, and perceptual quality across different methods, where circle size represents the number of model parameters required for end-to-end waveform inference. \textbf{(b)} Effect of the number of function evaluations (NFE) on perceptual quality and inference speed. NFE is defined as the number of neural-network evaluations performed during iterative sampling. Each curve represents a model, and each point is annotated with its corresponding NFE.}
\label{fig:efficiency_performance}
\end{figure*}

Additionally, Figure~\ref{fig:efficiency_performance} compares perceptual quality and computational efficiency from two complementary perspectives. As shown in Figure~\ref{fig:rtf_visqol_bubble}, AP-BWE and FLowHigh provide the strongest quality--efficiency trade-offs. AP-BWE achieves near-best perceptual quality with the lowest RTF, whereas FLowHigh obtains the highest ViSQOL while remaining substantially faster than the diffusion-based methods. In contrast, the discriminative models are computationally efficient but yield smaller perceptual improvements, while NU-Wave2 and UDM+ incur considerably higher inference costs without achieving better ViSQOL. 

Figure~\ref{fig:rtf_visqol_nfe} shows that increasing NFE does not necessarily yield proportional perceptual gains. FLowHigh maintains nearly unchanged ViSQOL from NFE 2 to 50, while its RTF increases from 0.0389 to 0.1596, since it is adapted to few-step generation. UniverSR improves substantially from NFE 2 to 10 but saturates thereafter, whereas the diffusion-based methods achieve more gradual improvements at considerably higher computational cost. NU-Wave2 even exhibits a slight decrease in perceptual quality at NFE 100. Overall, these results indicate that, in the two evaluated VCTK tasks, the selected flow-based methods, particularly FLowHigh, show favorable trade-offs between perceptual quality and inference efficiency.

\subsection{Discussion}
Discriminative \ac{dnn}-based methods formulate \ac{bwe}/\ac{asr} as a direct mapping from a \ac{bl}/\ac{lr} observation to a single \ac{bb}/\ac{hr} estimate~\citep{kuleshov2017audio,birnbaum2019temporal,rakotonirina2021self}. This formulation enables stable training and efficient inference but is less suited to the inherent ambiguity of missing \ac{hf} content. Point-estimation objectives may favor averaged predictions, resulting in spectral over-smoothing and reduced harmonic, transient, and aperiodic detail. This helps explain why discriminative methods can preserve the observed \ac{lf} content while providing more limited perceptual improvement.

Generative objectives can encourage sharper and more plausible \ac{hf} reconstruction, although not all generative methods explicitly represent conditional uncertainty. In particular, many GAN-based systems remain deterministic at inference despite using adversarial training to improve perceptual realism~\citep{mandel2023aero,lu2024towards,zhang2025vm}. Such objectives can produce detailed reconstructions but may also be unstable or generate content that is insufficiently constrained by the observed signal.

Diffusion, flow, and bridge-based methods introduce more explicit generation dynamics. Diffusion and score-based models learn iterative denoising processes that capture complex conditional distributions, but repeated neural-network evaluations increase inference cost~\citep{han2022nu,liu2024audiosr,yu2023conditioning}. Flow-based methods learn continuous transport between simple and target distributions, with recent approaches demonstrating the potential for efficient few-step sampling~\citep{yun2025flowhigh,welker2025real,hsieh2026towards}. Bridge-based methods connect degraded and clean distributions through stochastic paths, aligning naturally with restoration tasks while introducing additional path-design and sampling complexity~\citep{li2025bridge,li2025audio}.

Overall, these model families exhibit different trade-offs between reconstruction quality, inference efficiency, and modeling complexity rather than a universal performance ranking. We summarize their principal strengths, limitations, and inference characteristics in Table~\ref{tab:model_family_comparison}.

\begin{table*}[h]
\centering
\caption{\textbf{Comparison of representative model families for BWE/SR in terms of their strengths, limitations, and inference characteristics.}}
\label{tab:model_family_comparison}

\small
\renewcommand{\arraystretch}{1.3}
\setlength{\tabcolsep}{6.5pt}
\begin{tabular}{llll}
\toprule
\textbf{Model Family} & \textbf{Strengths} & \textbf{Limitations} & \textbf{Inference Characteristics} \\
\midrule
Discriminative
& Stable and fast
& Limited high-frequency detail
& Deterministic, single-pass \\

GAN
& Sharp reconstruction
& Unstable adversarial training
& Feed-forward, single-pass \\

Diffusion
& Strong distribution modeling
& High sampling cost
& Iterative, NFE-dependent \\

Flow
& High quality with few steps
& Sensitive to path design
& Few-step, solver-dependent \\

Bridge
& Restoration-aligned paths
& Complex path design
& Iterative, path-dependent \\
\bottomrule
\end{tabular}

\end{table*}

%% file: sec/8_discussion_final.tex
\section{Challenges and Outlook}
\label{sec:challenge}
This section discusses key challenges in perceptual evaluation, inference efficiency, robustness, and model design, and highlights future directions and broader impacts for \ac{bwe}/\ac{asr}.

\subsection{Challenges and Open Problems}
\subsubsection{Evaluation and Perceptual Quality}

Evaluating \ac{bwe}/\ac{asr} systems remains challenging because objective metrics do not always align with \ac{hf} reconstruction accuracy or perceived quality. PESQ, although widely used in \ac{se}, may be less suitable for \ac{bwe}/\ac{asr}, as higher scores do not necessarily indicate more accurate \ac{hf} recovery~\citep{hauret2023eben}. Similarly, lower LSD indicates greater spectral similarity but not necessarily more natural audio. STOI primarily measures intelligibility, while ViSQOL does not fully capture timbre, harmonic structure, transient fidelity, or the realism of generated \ac{hf} content.

This limitation is more pronounced for music and general audio, where quality depends on harmonic structure, transients, texture, and spatial characteristics. Subjective listening tests therefore remain essential for assessing naturalness, artifacts, and reconstruction plausibility, but they are costly and time-consuming. Reliable evaluation should combine signal-based, perceptual, and task-specific metrics with controlled listening tests. Developing reference-free metrics and standardized evaluation protocols that correlate more strongly with human judgments remains an important open problem.

\subsubsection{Inference Efficiency and Practical Deployment}
Inference efficiency is critical for practical \ac{bwe}/\ac{asr} systems, particularly in real-time and streaming applications. Single-pass models generally provide low latency, whereas iterative generative models require repeated neural-network evaluations and introduce a quality--efficiency trade-off. End-to-end runtime also depends on architectural complexity, feature extraction, and auxiliary components; therefore, parameter count or sampling steps alone cannot fully characterize efficiency.

Our analysis in Figure~\ref{fig:rtf_visqol_nfe} further shows that increasing NFE often raises RTF while yielding diminishing perceptual gains. Practical deployment should therefore jointly consider reconstruction quality, end-to-end RTF, latency, memory usage, throughput, and hardware settings. Causal or chunk-wise processing can reduce latency for long audio but may introduce boundary artifacts and weaken temporal consistency. Future evaluations should report these practical measures under clearly specified hardware and inference settings.

\subsubsection{Robustness and Model Design}
Robustness to real-world degradations remains limited because many \ac{bwe}/\ac{asr} systems are trained using simplified low-pass filtering or idealized downsampling operators. In practice, bandwidth-limited audio may also contain codec artifacts, background noise, reverberation, device coloration, nonlinear distortion, and unknown cutoff frequencies. Models trained on a narrow degradation distribution may therefore perform poorly under mismatched recording conditions, unseen domains, or combinations of degradations. Improving robustness requires more realistic degradation modeling, diverse training data, degradation-aware conditioning, and adaptation strategies for unknown input conditions.

The model design space also remains underexplored. Diffusion schedules, bridge endpoints, flow trajectories, conditioning mechanisms, and waveform, spectrogram, or latent representations can substantially affect reconstruction quality, diversity, and computational cost. Future work should investigate how these components interact and develop principled designs that balance fidelity to the observed \ac{lf} signal, plausible \ac{hf} generation, controllability, robustness, and inference efficiency.

\subsection{Future Outlook}
Recent significant advances in \acp{llm}~\citep{touvron2023llama,deepseekv3,nie2025large,wang2025reasoning,wu2025step,huo2026auden}, \acp{lalm}~\citep{wu2024just,Qwen2-Audio,chenaudio,ghosh2025audio,ghosh2026audioflamingonextnext,wu2025step,rong2026audiogenie,gong2026mossaudio,wang2026sightsound}, and multimodal pretraining~\citep{Qwen3-Omni,kong2025,wang2025reasoning,yao2025generality} suggest several promising directions for \ac{bwe}/\ac{asr}. First, large Transformer-based foundation models trained on audio-text or broader multimodal corpora can provide semantically structured priors over speech, music, general sounds, and complex acoustic scenes. Conditioning \ac{bwe}/\ac{asr} models on such high-level representations may help constrain \ac{hf} synthesis, reducing implausible hallucination while preserving perceptually meaningful detail.

Second, retrieval-augmented paradigms offer a possible route toward more controllable \ac{bwe}/\ac{asr}. Instead of relying solely on parametric model memory, future systems could retrieve high-quality exemplars matched by speaker identity, instrument family, acoustic scene, timbral profile, or caption-level semantics, and use these references to guide \ac{hf} reconstruction conditioned on the degraded input. Such retrieval mechanisms may be especially useful when the missing bandwidth depends strongly on source-specific structure, but they also raise open questions about retrieval quality, reference mismatch, and computational overhead.

Additionally, self-supervised and multimodal representation learning may improve robustness and evaluation. Objectives such as contrastive learning, masked audio modeling, and audio-language alignment can produce representations that transfer across restoration tasks and degradation conditions, potentially improving generalization to noise, codec artifacts, device coloration, and unseen bandwidth constraints. In parallel, learned perceptual metrics derived from audio-language or audio-visual models may better capture human judgments of naturalness, timbre, texture, and event realism than conventional signal-based measures, making them useful for evaluation, perceptual fine-tuning, and guided generation.

Overall, future progress in \ac{bwe}/\ac{asr} will likely depend not only on stronger generative backbones, but also on improved conditioning, retrieval, perceptual evaluation, and deployment-aware optimization. Joint advances across these directions will be essential for developing robust, efficient, and perceptually faithful systems across speech, music, and general audio.

\subsection{Broader Impact Statement}

\ac{bwe}/\ac{asr} technologies may synthesize plausible but inaccurate \ac{hf} details, speaker characteristics, or acoustic events that could be mistaken for authentic content in sensitive applications, including forensic analysis, surveillance, historical restoration, and accessibility tools. While these technologies can improve degraded recordings, they may also facilitate voice spoofing, impersonation, and deceptive media. Generated or enhanced audio should therefore be disclosed and supported by provenance metadata, watermarking, and human verification, particularly when used as evidence or for decision-making. Moreover, training datasets may contain identifiable voices, copyrighted recordings, and demographic or domain biases, raising privacy, licensing, and fairness concerns. Therefore, responsible development should protect data rights and privacy, evaluate potential biases, and ensure that synthetic or modified audio can be identified and traced.

\section{Literature Selection Process}
\label{sec:literature_selection}
We conducted a structured literature search of \ac{bwe}/\ac{asr} across Google Scholar, IEEE Xplore, ACM Digital Library, and arXiv. The covered publications are primarily from January 2015 to May 2026, while other related works fall outside this period. Search keywords included ``audio/speech/music super-resolution,'' ``audio/speech/music bandwidth extension,'' ``audio/speech/music restoration,'' and ``universal speech enhancement''. Additional relevant studies were identified through citation tracking, while seminal works published before 2015 were included to provide historical context.
We included studies addressing \ac{bwe}/\ac{asr} for speech, music, and general audio, spanning \ac{dl}-based discriminative and generative approaches. The scope of this survey is detailed in Section~\ref{subsec:survey_scope}. Both peer-reviewed publications and relevant preprints were considered to capture recent developments.
The selected studies were categorized according to their modeling paradigms and neural architectures. 
The final collection of searched works is now available in \href{https://github.com/ningyuan33/Awesome-Audio-Super-Resolution-Bandwidth-Extension}{\textcolor{blue}{Github}}\footnote{Repository at: \url{https://github.com/ningyuan33/Awesome-Audio-Super-Resolution-Bandwidth-Extension}.}.

%% file: sec/9_conclusion.tex
\section{Conclusion}
\label{sec:con}

This survey presented a unified review of \ac{bwe} and \ac{asr}, treating both as ill-posed \ac{hf} reconstruction problems while clarifying their relationship and distinctions. We traced the field's evolution from discriminative \ac{dnn}-based models to modern generative paradigms, including \ac{ar} models, \acp{vae}, \acp{gan}, diffusion and score-based models, flow-based models, and Schrödinger bridges. From this perspective, we highlighted a central transition in \ac{bwe}/\ac{asr}: from point estimation, which often leads to regression-to-the-mean behavior and spectral over-smoothing, to distribution-aware generation, which better captures the one-to-many nature of missing \ac{hf} content. We also conducted unified experiments on representative discriminative and generative methods, providing controlled empirical support for their comparative performance. We further summarized key trade-offs in reconstruction fidelity, perceptual quality, robustness, and computational efficiency, and identified open challenges in perceptual evaluation, scalable deployment, and real-world generalization. We hope this survey provides a clear taxonomy and practical roadmap for advancing robust, efficient, and perceptually faithful \ac{bwe}/\ac{asr} systems across speech, music, and general audio.

%% file: sec/10_acknowledgement.tex
\section*{Acknowledgments}
This work was supported in part by the U.S. Office of Naval Research under Grants N000142412004 and N000142312852.
The authors are grateful to Dr. Robin Scheibler from Google DeepMind for his valuable comments and expert feedback during the preparation of this manuscript. His suggestions helped to improve the organization of the survey and clarify several technical discussions.